\documentclass[useAMS,usegraphicx,usenatbib]{mn2e}

\usepackage{url}
\usepackage{times}

\usepackage{graphicx}
\usepackage{xcolor}

\usepackage{amssymb,amsmath}

\def\refjnl#1{{\rm#1}}%
\newcommand\aj{\refjnl{AJ}}%
\newcommand\apj{\refjnl{ApJ}}%
\newcommand\apjl{\refjnl{ApJ}}%
\newcommand\apjs{\refjnl{ApJS}}%
\newcommand\aap{\refjnl{A\&A}}%
\newcommand\mnras{\refjnl{MNRAS}}%
\newcommand\pasp{\refjnl{PASP}}%
\newcommand\nat{\refjnl{Nature}}%
\newcommand\nar{\refjnl{New~Astron.~Rev.}}%
\newcommand\na{\refjnl{New~Astron}}%

\newcommand{\acknowledgements}{\begin{small}\section*{Acknowledgments}\end{small}}
\newcommand{\tableline}{\hline}
\newcommand{\boldtext}{}

\title[Quenching of massive galaxies at high redshift as a result of cosmological starvation]{The Argo Simulation: I. Quenching of Massive Galaxies at High Redshift as a Result of Cosmological Starvation}

\author[R. Feldmann and L. Mayer]{Robert Feldmann,$^{1,\ast}$\thanks{Hubble fellow} and Lucio Mayer$^{2}$
\\
\normalsize{$^{1}$Department of Astronomy, University of California, Berkeley, CA 94720-3411, USA}\\
\normalsize{$^{2}$Center for Theoretical Astrophysics and Cosmology, 
Institute for Computational Science, University of Z\"{u}rich, 8057 Z\"{u}rich, Switzerland}\\
\\
\normalsize{$^\ast$Corresponding author. E-mail: feldmann@berkeley.edu}
}

\begin{document} 

\maketitle 

\begin{abstract}
Observations show a prevalence of high redshift galaxies with large stellar masses and predominantly passive stellar populations. 
A variety of processes have been suggested that could reduce the star formation in such galaxies to observed levels, including quasar mode feedback, 
virial shock heating, or galactic winds driven by stellar feedback. However, the main quenching mechanisms have yet to be identified.
Here we study the origin of star formation quenching using Argo, a cosmological, \boldtext{hydrodynamical} zoom-in simulation 
that follows the evolution of a massive galaxy at $z\geq{}2$. This simulation adopts the same sub-grid recipes of
the Eris simulations, which have been shown to form realistic disk galaxies, and, in one version, adopts also a mass
and spatial resolution identical to Eris. The resulting galaxy has properties consistent 
with those of observed, massive ($M_*\sim{}10^{11}$ $M_\odot$) galaxies at $z\sim{}2$ and with abundance matching predictions. 
Our models do not include AGN feedback indicating that supermassive black holes likely play a subordinate role in determining masses 
and sizes of massive galaxies at high $z$. The specific star formation rate (sSFR) of the simulated galaxy matches the 
observed $M_*$ - sSFR relation at early times. This period of smooth stellar mass growth comes to a sudden halt at $z=3.5$ when 
the sSFR drops by almost an order of magnitude within a few hundred Myr. 
The suppression is initiated by a leveling off and a subsequent reduction of the cool gas accretion rate onto the galaxy, and not by feedback processes. 
This ``cosmological starvation'' occurs as the parent dark matter halo switches from a fast collapsing mode to a slow accretion mode. 
Additional mechanisms, such as perhaps radio mode feedback from an AGN, are needed to quench any residual star formation of the galaxy and 
to maintain a low sSFR until the present time.
\end{abstract}


\section{Introduction}
\label{sect:Intro}
A large number of massive galaxies at $z\sim{}2$ are dominated by a passively evolving stellar population with little ongoing star formation (e.g., \citealt{2003ApJ...587L..79F, 2004Natur.430..184C, 2004ApJ...616...40F, 2005ApJ...626..680D, 2006ApJ...638L..59V, 2007A&A...476..137A, 2007ApJ...655...51W, 2008ApJ...682..896K, 2008A&A...482...21C, 2009ApJ...691.1879W, 2009ApJ...706L.173B, 2010ApJ...708..202M, 2010ApJ...709..644I, 2012ApJ...755...26O, 2013ApJ...764L...8B, 2013ApJ...770L..39W, 2014ApJ...780...34L, 2014A&A...561A..86M}).

The fraction of such galaxies is 20-30\% for galaxies with stellar masses $\sim{}2-5\times{}10^{10}$ $M_\odot$ and raises to $\sim{}50\%$ for galaxies with masses above $\sim{}10^{11}$ $M_\odot$ (e.g., \citealt{2011ApJ...735...86W, 2011ApJ...739...24B, 2013ApJ...777...18M, 2014arXiv1401.2984T}).   \boldtext{These galaxies are called quiescent, a reference to their weak or absent star formation activity, although the actual operational definition is typically based on U-V and V-J restframe colors, see section \ref{sect:CompObs}}. So far, no consensus has been reached on the nature of the physical processes responsible for suppressing (``quenching'') star formation in massive galaxies at high redshift.

Empirical studies show that out to at least $z\sim{}1$ the likelihood of a galaxy being quenched is correlated both with its stellar mass (``mass quenching'') and with the environmental density of neighboring galaxies (``environmental quenching''), see \cite{2010ApJ...721..193P}. The latter quenching channel is primarily affecting satellite galaxies (e.g., \citealt{2012ApJ...757....4P, 2013MNRAS.432..336W}), while the quiescent fraction of central galaxies appears to be not strongly correlated with current environmental overdensity (\citealt{2014MNRAS.438..717K} and references therein). Environmental quenching is likely related to processes preferentially occurring in galaxy groups and clusters, such as ram-pressure stripping of the interstellar medium (ISM; \citealt{1972ApJ...176....1G, 1999MNRAS.308..947A}), the reduction of gas accretion onto satellites (``starvation'', \citealt{1980ApJ...237..692L, 2000ApJ...540..113B, 2008ApJ...672L.103K, 2008MNRAS.383..593M, 2008MNRAS.387...79V, 2011ApJ...736...88F, 2013MNRAS.430.3017B}), or frequent galaxy interactions (``harassment'', \citealt{1981ApJ...243...32F, 1996Natur.379..613M}).

Most mass quenching mechanisms discussed in the literature either prevent or decrease the accretion of gas onto galaxies or eject gas from galaxies. For instance, the growth of a stable virial shock in massive halos reduces the accretion of relatively cool, not strongly shock-heated, gas onto galaxies \citep{2003MNRAS.345..349B, 2005MNRAS.363....2K, 2006MNRAS.368....2D, 2006MNRAS.370.1651C}. In addition, cooling of shock-heated gas might be counteracted by gravitational heating from infalling satellite galaxies or gas clumps \citep{2008MNRAS.383..119D, 2008ApJ...680...54K, 2009ApJ...697L..38J, 2012ApJ...754..115J}.

Other quenching channels are related to active galactic nuclei (AGN) and to star formation in galaxies. Major mergers of gas-rich disk galaxies at high redshift may ignite powerful starbursts \citep{1986ApJ...303...39D, 2005ApJ...618..569M} and result in quasar-activity \citep{1988ApJ...325...74S, 2004ApJ...608...62S, 2005MNRAS.361..776S, 2005ApJ...620L..79S, 2005Natur.433..604D, 2006ApJS..163....1H, 2011MNRAS.412.1965M} that may launch outflows and remove large amounts of gas from galaxies. AGNs can also operate in a jet-powered \emph{radio mode} in which they counteract the cooling of gas from the hot halo surrounding galaxies with low intensity mechanical heating (e.g., \citealt{2006MNRAS.365...11C, 2006MNRAS.370..645B, 2007MNRAS.380..877S}).

Ejective feedback originating in quasar activity could explain why the entropy in galaxy clusters is close to the value needed to offset gas cooling \citep{2004ApJ...608...62S}. It also offers a physical basis for the co-evolution of black-hole mass and the bulge mass of the galaxy host \citep{1998AJ....115.2285M, 2003ApJ...589L..21M, 2004ApJ...604L..89H} and for the correlation between quasar activity and star formation \citep{1998MNRAS.293L..49B, 2003MNRAS.346.1055K}. However, it is still an open question whether quasar driven outflows triggered by galaxy interactions or mergers actually quench star formation in massive, high redshift galaxies (see \citealt{2012NewAR..56...93A} for a recent review). In fact, at high redshift there is little evidence for a link between merging and quasar activity \citep{2011ApJ...727L..31S, 2012ApJ...744..148K}, except at the highest quasar luminosities \citep{2012ApJ...758L..39T}, or between quasar activity and star formation \citep{2014ApJ...783...40G}.

Galaxy evolution simulations that include quasar mode feedback have not yet converged on a conclusive answer either, primarily because there exists a variety of choices as to which physical processes are modeled and how they are implemented (e.g., \citealt{2010ApJ...722..642O, 2010MNRAS.406L..55D, 2011ApJ...737...26N, 2011MNRAS.412..269P, 2012MNRAS.425..605F, 2013MNRAS.434.3606N, 2013MNRAS.431.2513W, 2014arXiv1402.4482G}).

Numerical simulations agree, however, that radio mode feedback can reduce the $z=0$ stellar masses of central galaxies in clusters, that it can result in more realistic galaxy colors and also better agreement with observed X-ray scaling relations \citep{2006MNRAS.366..397S, 2008ApJ...687L..53P, 2008MNRAS.387...13K, 2010MNRAS.406..822M, 2010MNRAS.409..985D, 2011MNRAS.417.1853D, 2011MNRAS.414..195T, 2013MNRAS.433.3297D, 2013MNRAS.436.1750R, 2014MNRAS.438..195P}. However, this feedback channel primarily affects the hot gas halo surrounding galaxies, and typically does not launch large scale winds involving the ISM of the galaxy. Hence, radio mode feedback appears to be an unlikely candidate for triggering a rapid star formation quenching in high redshift galaxies, but it may play an important role in maintaining low star formation activity in already quenched galaxies (e.g., \citealt{2005MNRAS.362...25B, 2006MNRAS.365...11C}). 

Hence, we are left with two related, but distinct, questions. First, why do high redshift galaxies leave the star forming sequence \citep{2007ApJ...660L..43N, 2007A&A...468...33E, 2007ApJ...670..156D, 2011A&A...533A.119E}? We will show that massive, high redshift galaxies can only remain on the star forming sequence for so long before the finite supply of gas limits their star formation activity. Specifically, during an early collapse phase, gas accretion rates and sSFRs are high and massive galaxies form and assemble much of their stellar mass. This is followed by a cosmological starvation phase in which accretion rates, and subsequently star formation activity, level off and eventually decline. Second, why do galaxies shut-down their star formation almost completely instead of maintaining a low, but non-negligible star formation activity? We will argue that a combination of stellar feedback and, potentially, AGN radio mode feedback, coupled with gravitational heating is required to complete the transition from the star forming sequence to the quiescent galaxy population. 

\begin{table*}
\begin{center}
\begin{tabular}{cccccccc}
\tableline
Run & $m_{\rm gas}$             & $m_{\rm star}$           & $m_{\rm DM}$           & $\epsilon_{\rm bar}$ & $\epsilon_{\rm DM}$ & $n_{\rm SF}$ & $z_{\rm end}$ \\ 
                  & ($M_\odot$ $h^{-1}$)  & ($M_\odot$ $h^{-1}$) & ($M_\odot$ $h^{-1}$) & (pc $h^{-1}$)             & (pc $h^{-1}$) & (m$_{\rm H}$ cm$^{-3}$) \\ \tableline
LR*             & $9.9\times{}10^5$        & $2.9\times{}10^5$     & $4.7\times{}10^6$       & 219                           &  365 & 0.1 & 1 \\
MR1          & $1.2\times{}10^5$        & $3.7\times{}10^4$     & $5.8\times{}10^5$       & 109                           & 183 & 0.1 & 2 \\
MR2          & $1.2\times{}10^5$        & $3.7\times{}10^4$     & $5.8\times{}10^5$       & 109                           & 183 & 5 & 2 \\
HR            & $1.5\times{}10^4$        & $4.6\times{}10^3$     & $5.8\times{}10^5$       &  88                            & 183 & 5 & 3.4 \\
\tableline
\end{tabular}
\caption{Overview of the individual runs performed as part of the Argo project. The first four columns provide the label and the masses of gas, star, and DM particles in the zoom-in region of the run. The next two columns are the gravitational spline softening lengths of baryonic (i.e., gas and star) particles and of DM particles in proper units. The penultimate column shows the star formation threshold. The final column indicates the redshift at which the run is stopped. Runs MR1 and MR2 start from the same initial conditions and differ only in the adopted star formation threshold. \boldtext{*We carried out at this resolution: the default LR run with our fiducial physics model, the LR-noFB run without energetic feedback from supernovae, the LR-noFB(z$<$4) run without supernova feedback after $z=4$, and the LR-noML(z$<$3) run without stellar mass loss after $z=3$.}}
\label{tab:Res}
\end{center}
\end{table*}

Our results are based on a cosmological, zoom-in simulation of a massive, high redshift galaxy. The simulated galaxy resides at the center of a halo ($M_{\rm vir}\sim{}10^{13}$ $M_\odot$ at $z=0$) that should harbor common (i.e., of intermediate stellar mass $\sim{}1-3\times{}10^{11}$ $M_\odot$) quiescent galaxies in the local Universe. The halo is located in a typical, mildly over-dense region that does not contain a more massive halo. Hence, our findings likely apply to a majority of massive, quiescent, central galaxies.

Despite their importance and ubiquity, galaxies in $\sim{}10^{13}$ $M_\odot$ halos have only recently been targeted by cosmological, zoom-in simulations \citep{2008ApJ...672L.103K, 2010ApJ...709..218F, 2011ApJ...736...88F, 2010ApJ...725.2312O, 2012ApJ...744...63O, 2014ApJ...781...38C}. One of the reasons is the numerical challenge of resolving galaxies in such massive halos. In addition, physical models are often tuned based on simulations of lower mass, star forming galaxies (e.g., \citealt{2006MNRAS.373.1074S}). Hence, it is not obvious that the same models are adequate to simulate massive, quiescent galaxies. Fortunately, we can gauge the realism of our numerical approach even before starting the Argo simulation. In particular, simulations run with the same code, with similar methodology, and at a comparable resolution produce dwarf galaxies \citep{2013arXiv1308.4131S}, Milky-Way like galaxies \boldtext{(e.g., the Eris simulation, \citealt{2011ApJ...742...76G})}, and massive galaxies \citep{2010ApJ...709..218F} with reasonably realistic properties. The same numerical approach is also able to reproduce the enrichment level of the circum-galactic gas around high redshift galaxies \citep{2013ApJ...765...89S}.

The paper is organized as follows. We outline the set-up of the Argo simulation and the strategy of the data analysis in section \ref{sect:Sim}. We present our first results in section \ref{sect:CompObs} where we compare the properties of the simulated galaxy with those of observed high redshift galaxies. In section \ref{sect:LeavingSFS} we then analyze the star formation history of the simulated galaxy finding an onset of quenching at $z\sim{}3.5$. We identify cosmological starvation as the cause of the star formation suppression in section \ref{section:OriginQuenching}. We highlight potential caveats that may affect the conclusion drawn from our work in section \ref{sect:Caveats}. We compare our results to previous theoretical works in section \ref{sect:Discussion}. Finally, in section \ref{sect:Summary} we end this paper with a summary of our main findings and our conclusions.

\section{Details of the Simulation and the Data Analysis}
\label{sect:Sim}

In this paper we present Argo, a cosmological, hydrodynamical simulation of a group-sized dark matter halo and its gaseous and stellar content. The virial mass of the selected dark matter halo is $\sim{}3\times{}10^{12}$ $M_\odot$ at $z=2$ and $\sim{}2\times{}10^{13}$ $M_\odot$ at $z=0$. Here and throughout the paper, the virial mass of a halo refers to the mass enclosed within a spherical volume centered on the density peak of the halo with a mean matter density of 180 times the background density of the Universe. 

The Argo simulation consists of a set of individual runs, with different choices for the numerical resolution or model parameters, of the same physical system. The four main runs (LR, MR1, MR2, and HR) of the Argo project are summarized in Table \ref{tab:Res}. The simulation is a follow-up of the $G2$ simulation introduced in \cite{2010ApJ...709..218F}. However, some of the runs have significantly improved numerical resolution and updated sub-grid model parameters that match those of the ``Eris'' simulation \citep{2011ApJ...742...76G}. 

We select the dark matter halo with a target $z=0$ virial mass from a 512$^3$ N-body simulation of a (123 Mpc)$^3$ cosmological volume \citep{2007MNRAS.375..489H}. We ensure that there is not a more massive halo within 7 Mpc and that the selected halo is not sitting in a void. The average overdensity is 1.4 when measured within in a sphere of 7 Mpc radius centered on the selected halo. No additional selection criteria are used.

To increase the numerical resolution at and around the halo we identify the Lagrangian patch in the initial conditions ($z_{\rm init}=41.5$) that contains all particles that enter a sphere of radius $R(z)=2\times{}R_{\rm vir}(z=0)/(1+z)$ around the most massive progenitor of the halo at any $z\geq{}z_{\rm m}$. We use $z_{\rm m}=2$ for the runs MR1, MR2, and HR and $z_{\rm m}=0$ for run LR. We use grafic-2 \citep{2001ApJS..137....1B} to add higher frequency density fluctuations to the original dark matter simulation and to obtain dark matter and SPH particle positions and velocities in the Zeldovich approximation. We refine the Lagrangian patch to the chosen dark matter resolution, see Table \ref{tab:Res}, and embed it into spherical shells of decreasing resolution. For each DM particle in the high resolution Lagrangian patch of the LR and MR runs we add one SPH particle while ensuring the proper baryonic power spectrum of the gas particles. We create initial conditions for the HR run based on the initial condition of the MR runs using a basic SPH particle splitting scheme. In particular, we split each SPH particle into 8 lower-mass SPH particles and offset their positions relative to the position of the original SPH particle.

The initial power spectrum of density fluctuations is generated with linger \citep{1995astro.ph..6070B} and is compatible with Wilkinson Microwave Anisotropy Probe-3 cosmological parameters \citep{2007ApJS..170..377S}. Specifically, we use $\Omega_{\rm m}=0.24$, $\Omega_\Lambda=0.76$, $\Omega_{\rm b}=0.04185$, $H_{\rm 0}=73$ km s$^{-1}$ Mpc$^{-1}$, $\sigma_8=0.77$, and $n=0.96$.

The Argo simulation is run with the parallel TreeSPH code GASOLINE \citep{2004NewA....9..137W}. GASOLINE is based on the parallel, multiple time stepping N-body code PKDGrav \citep{2001PhDT........21S}. We keep the gravitational softening length fixed in proper coordinates for $z<9$ and fixed in comoving coordinates for $z\geq{}9$. We provide the mass and force resolution of the various runs in Table \ref{tab:Res}. 

In addition to gravitational and hydrodynamical processes, the Argo simulation includes heating caused by a spatially uniform, redshift dependent UV radiation background \citep{1996ApJ...461...20H, 2011ApJ...742...76G}, (optically thin) radiative cooling for a primordial gas composition \citep{2004NewA....9..137W}, and sub-grid models for star formation and stellar feedback \citep{2006MNRAS.373.1074S}. Star particles (representing single stellar populations) form in a probabilistic fashion out of gas that is part of a convergent flow, has temperatures below $3\times{}10^4$ K, and spatial densities above $n_{\rm SF}$, see Table \ref{tab:Res}. The SFR proceeds at a rate $\dot{\rho}_*=\epsilon_{\rm SF}\rho_{\rm g}/t_{\rm dyn}\propto{}\rho_{\rm g}^{1.5}$. The star formation efficiency per free-fall time, $\epsilon_{\rm SF}$, is 5\%. Sub-grid stellar winds return a moderate fraction ($\sim{}40\%$ within a few Gyr) of the mass bound in star particles back to the gas. 

The number of supernovae type II and Ia that occur during a given time step is computed based on a \cite{1993MNRAS.262..545K} initial stellar mass function (IMF) and the stellar tracks by \cite{1996A&A...315..105R}. \boldtext{Each supernova injects metals and a thermal energy of} $8\times{}10^{50}$ erg into the neighboring SPH particles. We use the analytic blastwave scenario of \cite{1977ApJ...218..148M} to estimate the radius and cooling time of the unresolved supernova type II blast waves. We suspend cooling within the SPH kernel surrounding the star particle for a time corresponding to the end of the snowplow phase of the supernova type II blastwave. For typical ISM conditions the shutoff-time is of the order of a few $\sim{}10^{5}-10^{6}$ yr. Cooling is not suspended for supernovae of type Ia.

\begin{figure*}
\begin{tabular}{cc}
\includegraphics[width=80mm]{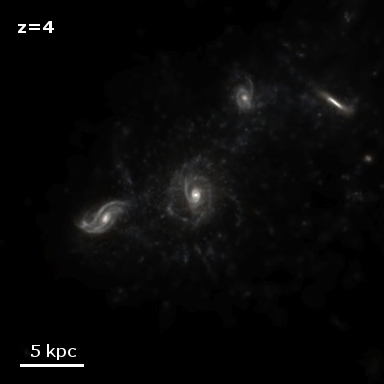} & 
\includegraphics[width=80mm]{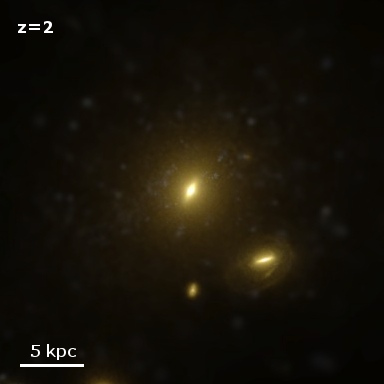} \\
\end{tabular}
\caption{Evolution of the simulated massive galaxy (center) over the redshift range z=4 (left panel, HR run) to z=2 (right panel, MR2 run). The panels show composite mock images in the HST I, J and H bands (logarithmic stretch). The size indicated at the bottom left is in proper kpc. Between $z=4$ and $z=2$ the massive galaxy evolves from a blue, compact, star forming galaxy with a disk component into a much redder, still compact, early-type galaxy.}
\label{fig:image}
\end{figure*}
 
The low resolution run (LR) is continued to $z=1$, the medium resolution runs (MR1, MR2) to $z=2$ and the high resolution run (HR) down to $z=3.4$. We store snapshots every $\sim{}70-150$ Myr.

We identify dark matter halos and sub-halos with the AMIGA Halo Finder (AHF; \citealt{2004MNRAS.351..399G, 2009ApJS..182..608K}). We run AHF on the total matter field generated by star, gas, and dark matter particles. In this paper we focus on the most massive galaxy in the high resolution region and on its dark matter halo. We use the merger tree tool provided by AHF to trace this galaxy from one snapshot to the next. In addition, we use the merger tree tool to identify sub-halos/satellite galaxies surrounding the main galaxy. Sometimes we find it useful to remove these sub-halos/satellite galaxies in our analysis, see below. We found that AHF sometimes identifies substructures within the main galaxy that are, e.g., star forming regions or density enhancements in spiral arms. Hence, we exclude substructures within a 2 kpc physical radius from the center of the galaxy from our substructure list. 

Reported stellar masses, optical and infrared magnitudes, and star formation rates of the simulated galaxy do not include contributions from satellite galaxies and are measured within a spherical radius of 12 kpc, unless stated otherwise. We checked that increasing the radius to 20 kpc changes these properties only at the few percent level. This is probably not surprising given that the half mass radius of the central galaxy is $\lesssim{}1$ kpc at $z\geq{}2$, see section \ref{sect:CompObs}. For the $z=1$ snapshot (LR run only) we use a somewhat larger radius of 20 kpc to account for the increased size of the galaxy at late times. Half mass radii are 2-dimensional apertures including exactly half the stellar mass (as defined above) of a galaxy. As projected radii depend on viewing direction (typical variations are of the order of 10\%), we quote mean half-mass radii averaged over 25 different lines of sight, including 20 random lines of sight, the $x$, $y$, and $z$ projection, and a face-on and an edge-on projection. Star formation rates at any given time refer to the stellar mass formed within the past 100 Myr, unless otherwise noted.

We use the stellar population synthesis model of \cite{2003MNRAS.344.1000B} to estimate the luminosities associated with each individual stellar particle. In particular, the model interprets each stellar particle as a single stellar population with a \cite{2003PASP..115..763C} IMF and with the mass, age, and metallicity as recorded by the stellar particle. Whenever necessary we interpolate logarithmically between the provided grid of metallicities and stellar population ages. We compute both rest frame magnitudes (in the AB system) in standard Bessel U and V filter bands and in the NSFCam J filter band\footnote{More specifically, J refers to the NSFCam ``Mauna Kea'' J-MK filter, see \url{http://irtfweb.ifa.hawaii.edu/~nsfcam2/Filter_Profiles.html}.}, as well as apparent K-band\footnote{We chose the WFCam K-band filter used in the UKIRT Infrared Deep Sky Survey, see Hewitt et al. 2006 and \url{http://www.ukidss.org/technical/photom/photom.html}.} magnitudes. We add a moderate amount of dust extinction with $A_V=0.5$ mag to account for the non-negligible amount of extinction in quiescent galaxies observed at $z\geq{}2$ (e.g., \citealt{2013ApJ...770L..39W, 2014arXiv1402.0003M}). We adopt the dust extinction curve described by \cite{2000ApJ...533..682C}.

Gas accretion rates onto the central galaxy are measured in spherical shells using the velocities and masses of gas particles. The bulk radial velocity of accreting or outflowing gas $v$ is derived from the gas accretion rate $\dot{M}$, the width of the radial shell $\Delta{}R$ and the mass within the shell $M$ as follows: $v = \Delta{}R\, \dot{M} / M$. Accretion rates and bulk velocities are corrected for the background Hubble flow.

\section{Comparison with Observations}
\label{sect:CompObs}

In this section we demonstrate that many properties of the simulated massive galaxy evolve strongly over the redshift range $z=2-4$. In addition, we show that by $z=2$ stellar masses, sizes, and colors are broadly consistent with those of massive, quiescent galaxies observed at those redshifts.

The morphology of the simulated massive galaxy changes substantially in the $\sim{}1.8$ Gyr interval between $z=4$ and $z=2$, see Figure~\ref{fig:image}. In particular, the galaxy evolves from a moderately massive ($M_*\sim{}10^{10}$ $M_\odot$), compact ($R_{1/2}\sim{}0.3$ kpc), blue ($U-V\sim{}0.7$) galaxy with a disk component into a still compact ($R_{1/2}\sim{}1$ kpc), massive ($M_*\sim{}10^{11}$ $M_\odot$), early-type galaxy with significantly redder colors ($U-V\sim{}1.25$). We summarize the evolution of many global properties of the simulated galaxy in Table~\ref{tab:Results}. 

\begin{table*}
\begin{center}
\begin{tabular}{ccccccccccccc}
\tableline
z & Run            & $\log$ $M_{\rm vir}$   & $\log$ $M_*$   &  $\log$ $M_{\rm cgas}$ & $R_{\rm vir}$ & $R_{1/2}$ &  $\log$ sSFR & $\log$ sSFR' & K$_{\rm obs}$ &  U$_{\rm rest}$ & V$_{\rm rest}$ & J$_{\rm rest}$ \\
   &              & ($M_\odot$)  & ($M_\odot$) & ($M_\odot$) & (kpc) & (kpc) & (yr$^{-1}$) & (yr$^{-1}$)  & mag$_{\rm AB}$ & mag$_{\rm AB}$ & mag$_{\rm AB}$ & mag$_{\rm AB}$ \\ \tableline
5 & LR & 11.02 & 9.27 & 9.51 & 26.2 & 0.24 & -8.05 & -8.12 & 26.1 & -20.4 & -20.8 & -20.8 \\
5 & MR1 & 10.98 & 9.33 & 9.49 & 25.4 & 0.26 & -8.58 & -8.32 & 26.5 & -19.9 & -20.5 & -20.6 \\
5 & MR2 & 11.03 & 9.47 & 9.34 & 26.4 & 0.20 & -8.54 & -8.84 & 26.3 & -20.1 & -20.7 & -20.9 \\
5 & HR & 10.97 & 9.31 & 9.45 & 25.1 & 0.21 & -8.44 & -8.65 & 26.6 & -19.9 & -20.4 & -20.6 \\
\tableline
4 & LR & 11.83 & 10.17 & 9.87 & 59.0 & 0.28 & -8.77 & -8.84 & 24.1 & -21.4 & -22.1 & -22.4 \\
4 & MR1 & 11.82 & 10.19 & 9.77 & 58.6 & 0.36 & -8.83 & -8.94 & 24.0 & -21.4 & -22.2 & -22.4 \\
4 & MR2 & 11.82 & 10.27 & 9.86 & 58.7 & 0.26 & -8.84 & -8.79 & 23.9 & -21.5 & -22.3 & -22.6 \\
4 & HR & 11.82 & 10.00 & 9.82 & 58.4 & 0.33 & -8.59 & -8.47 & 24.4 & -21.1 & -21.7 & -21.9 \\
\tableline
3.5 & LR & 12.03 & 10.73 & 9.83 & 76.4 & 0.51 & -8.66 & -8.85 & 22.3 & -22.9 & -23.6 & -23.9 \\
3.5 & MR1 & 12.03 & 10.65 & 9.76 & 76.5 & 0.55 & -8.58 & -9.01 & 22.5 & -22.8 & -23.4 & -23.7 \\
3.5 & MR2 & 12.03 & 10.72 & 9.85 & 76.5 & 0.39 & -8.76 & -8.96 & 22.4 & -22.7 & -23.4 & -23.7 \\
3.5 & HR & 12.03 & 10.59 & 9.67 & 76.3 & 1.00 & -8.74 & -8.87 & 22.8 & -22.4 & -23.1 & -23.4 \\
\tableline
3.4 & LR & 12.05 & 10.76 & 9.74 & 79.4 & 0.43 & -8.98 & -9.49 & 22.4 & -22.7 & -23.5 & -23.8 \\
3.4 & MR1 & 12.05 & 10.67 & 9.52 & 79.5 & 0.51 & -8.99 & -9.43 & 22.6 & -22.4 & -23.2 & -23.5 \\
3.4 & MR2 & 12.05 & 10.74 & 9.67 & 79.5 & 0.34 & -9.16 & -9.38 & 22.5 & -22.5 & -23.3 & -23.7 \\
3.4 & HR & 12.05 & 10.59 & 9.52 & 79.1 & 0.83 & -9.13 & -9.22 & 22.9 & -22.1 & -22.9 & -23.2 \\
\tableline
3 & LR & 12.10 & 10.82 & 9.55 & 90.1 & 0.43 & -9.26 & -9.85 & 22.3 & -22.3 & -23.3 & -23.7 \\
3 & MR1 & 12.12 & 10.71 & 9.34 & 91.6 & 0.56 & -9.38 & -9.85 & 22.7 & -21.9 & -22.9 & -23.3 \\
3 & MR2 & 12.12 & 10.78 & 9.61 & 91.5 & 0.44 & -9.37 & -9.67 & 22.5 & -22.0 & -23.1 & -23.5 \\
\tableline
2 & LR & 12.45 & 11.01 & 9.64 & 154.0 & 0.96 & -9.65 & -9.78 & 21.5 & -21.8 & -23.0 & -23.7 \\
2 & MR1 & 12.47 & 10.89 & 9.33 & 156.8 & 0.84 & -9.73 & -9.86 & 21.8 & -21.4 & -22.7 & -23.4 \\
2 & MR2 & 12.47 & 10.96 & 9.32 & 156.2 & 0.77 & -9.66 & -10.27 & 21.6 & -21.7 & -22.9 & -23.6 \\
\tableline
1 & LR & 12.86 & 11.29 & 9.39 & 323.7 & 1.71 & -10.22 & -10.60 & 19.6 & -21.4 & -22.9 & -23.8 \\
\tableline
\end{tabular}
\caption{Properties of the simulated central galaxy at selected redshifts. The first two columns label redshift and run, respectively. The remaining columns provide: (3) virial mass of the halo (the mean halo density is 180 times the background density of the Universe), (4) stellar mass within a 12 kpc radius (for $z\geq{}2$) or a 20 kpc radius (for $z=1$), (5) mass of gas colder than $T=3\times{}10^4$ K within a 12 kpc radius ($z\geq{}2$) or a 20 kpc radius ($z=1$), (6) viral radius of the halo, (7) average projected radius containing half the stellar mass given in column (4), (8) sSFR within 12 kpc (for $z\geq{}2$) or 20 kpc (for $z=1$) averaged over the past 20 Myr, (9) same as column (8), but excluding star formation and stellar mass within the central 0.3 kpc, (10) WFCAM K band magnitude in the observed frame within 12 kpc (for $z\geq{}2$) or 20 kpc (for $z=1$), (11-12) rest frame Bessel U and V band magnitudes, and (13) rest frame NSFCAM J band magnitude. Magnitudes are given in the AB system and are derived based on the age, mass, and metallicity of each stellar particle using \protect\cite{2003MNRAS.344.1000B} stellar synthesis models and a \protect\cite{2003PASP..115..763C} IMF. We apply a uniform \protect\cite{2000ApJ...533..682C} dust screen of A$_{V}=0.5$ mag to account for a moderate amount of extinction observed in many quiescent high redshift galaxies \protect\citep{2014arXiv1402.0003M}. We identify and mask any satellite galaxy before measuring the properties shown in columns 4 and 5 and in columns 7 through 13.}
\label{tab:Results}
\end{center}
\end{table*}

Galaxies show a bimodal distributions of galaxy colors and star formation rates (SFRs) both in the local Universe (e.g., \citealt{2003MNRAS.341...33K, 2003ApJ...594..186B, 2004ApJ...601L..29H, 2004ApJ...615L.101B}) and out to at least  $z\sim{}2$ (e.g., \citealt{2004ApJ...608..752B, 2009ApJ...694.1171T, 2009ApJ...691.1879W, 2009ApJ...706L.173B, 2011ApJ...739...24B, 2011ApJ...735...86W, 2013ApJ...777...18M}). Consequently, empirical cuts in color-color space or thresholds in sSFR are often used to classify galaxies as ``quiescent'' or ``star forming''. A popular method of the first kind is a classification based on the U-V and V-J rest frame colors, the so-called UVJ diagram (e.g., \citealt{2005ApJ...624L..81L, 2007ApJ...655...51W, 2009ApJ...691.1879W, 2013ApJ...771...85V}), which employs the fact that galaxy light dominated by an old stellar population shows a pronounced 4000\AA{} break in its spectrum. Cuts in sSFR are often based on a threshold of ${\rm sSFR}(z)=0.3/t_{\rm H}$, where $t_{\rm H}$ is the age of the universe at the given redshift \citep{2008ApJ...688..770F, 2009A&A...501...15F, 2010ApJ...713..738W}. This corresponds to $\sim{}10^{-10}$ yr$^{-1}$ at $z=2-3$. 

It should be kept in mind that the classification is based on the observed bimodality and not on matching galaxy properties to those of local star forming or quiescent galaxies. In fact, galaxy properties change dramatically between $z\sim{}2$ and today. For instance, star forming galaxies at high $z$ have larger gas fraction \citep{2008ApJ...673L..21D, 2010Natur.463..781T, 2013ApJ...768...74T, 2010ApJ...713..686D, 2012MNRAS.421...98D, 2012ApJ...760....6M, 2013MNRAS.433.1910F, 2013arXiv1303.4392S}, lower metallicities at a given stellar mass \citep{2006ApJ...644..813E, 2008A&A...488..463M, 2010MNRAS.408.2115M}, and a more turbulent interstellar medium (ISM; \citealt{2006ApJ...645.1062F}). High redshift galaxies that are quiescent (based on the UVJ diagram) have younger stellar populations than local quiescent galaxies (e.g., \citealt{2013ApJ...770L..39W}), significant dust extinction (e.g., \citealt{2014arXiv1402.0003M}) and may also be forming stars at a substantial rate (e.g., \citealt{2011ApJ...739...24B, 2014ApJ...783L..14S}, but cf. \citealt{2014ApJ...783L..30U}).

\begin{figure}
\begin{tabular}{c}
\includegraphics[width=77mm]{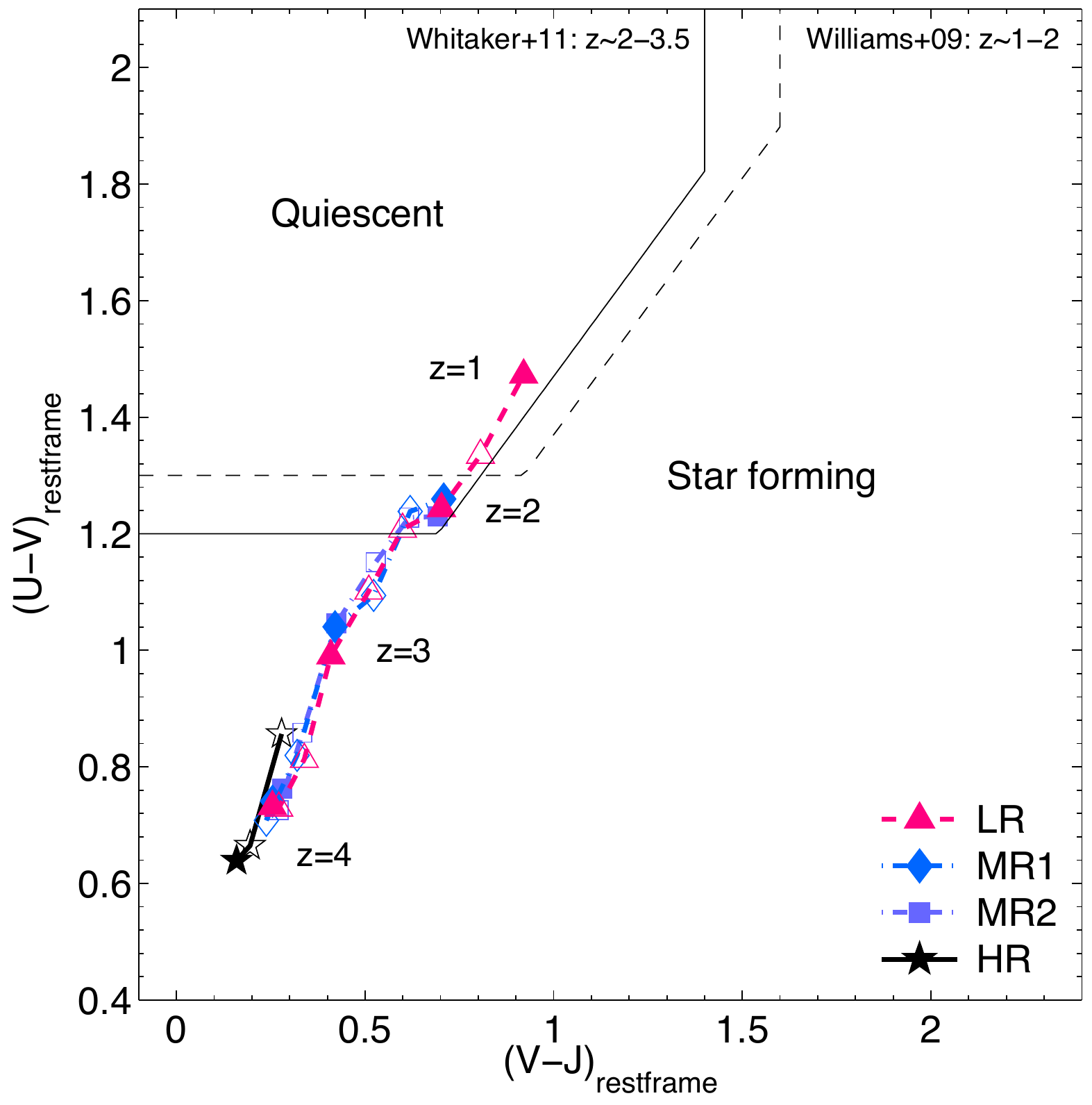}
\end{tabular}
\caption{Restframe UVJ colors of the simulated galaxy from $z=4$ to $z=1$. Symbols connected with solid, dot-dashed, and dashed lines show the U-V and V-J colors of the central galaxy in the HR, MR, and LR runs (see legend). Filled symbols mark the specific redshifts 4, 3, 2, and 1 (from bottom left to top right). Observed galaxies are often classified as quiescent or star forming based on their position in the UVJ diagram. We show two different selection choices: one for galaxies at  $z\sim{}2-3.5$ (solid black line; \citealt{2011ApJ...735...86W}) and one for galaxies at $z\sim{}1-2$ (dashed black line; \citealt{2009ApJ...691.1879W}), see legend. At $z\lesssim{}2$ the simulated galaxy would likely be classified as a quiescent galaxy based on its location in the UVJ diagram.}
\label{fig:UVJ}
\end{figure}

We explore the quiescent or star forming nature of the simulated galaxy in the UVJ diagram of Figure~\ref{fig:UVJ}. The galaxy starts with blue U-V and V-J colors and, with time, turns redder in both colors. At $z\sim{}2$ the galaxy crosses over into a region in color-color space that is associated with quiescent high redshift galaxies. At this time its sSFR falls below $\sim{}10^{-10}$ yr$^{-1}$, i.e., below the sSFR threshold mentioned above. Note, however, that the actual SFRs are still of the order of $\sim{}10$ $M_\odot$. Hence, although the galaxy qualifies as quiescent according to both the color-color and the sSFR criteria, it differs substantially from completely ``read and dead'' galaxies in the local Universe. Reassuringly, the results for $z=1$, see Figure~\ref{fig:UVJ} and Table~\ref{tab:Results}, show that the sSFR continues to decrease and the colors continue to redden with increasing cosmic time. We note that there is good agreement among the various runs.

A problem in many numerical simulations is that they tend to produce galaxies that are overly massive for their parent dark matter halo. This is a problem for Milky-Way sized halos (see \citealt{2012MNRAS.423.1726S} and reference therein), for central galaxies in groups (e.g., \citealt{2010ApJ...709..218F}), but especially for cluster central galaxies (e.g., \citealt{2011ASL.....4..204B}). Improved numerical resolution and modeling of stellar feedback has alleviated this problem substantially for galaxies in Milky-Way sized (and smaller) halos \citep{2011ApJ...742...76G, 2012MNRAS.424.1275B, 2013MNRAS.434.3142A,2013ApJ...766...56M, 2013arXiv1311.2073H} and AGN feedback offers a potential solution for galaxies in clusters (e.g., \citealt{2011MNRAS.414..195T, 2012MNRAS.420.2859M}). More specifically, an important success of the \cite{2011ApJ...742...76G} model is that the obtained stellar mass at $z=0$ is consistent with the observed stellar mass of the Milky-Way. However, a priori it is unclear, whether this model, which includes thermal feedback from supernovae as its only energetic feedback source, is adequate to prevent over-cooling in group sized halos, even at high redshift (see \citealt{2013MNRAS.428..129S}).

\begin{figure}
\begin{tabular}{c}
\includegraphics[width=80mm]{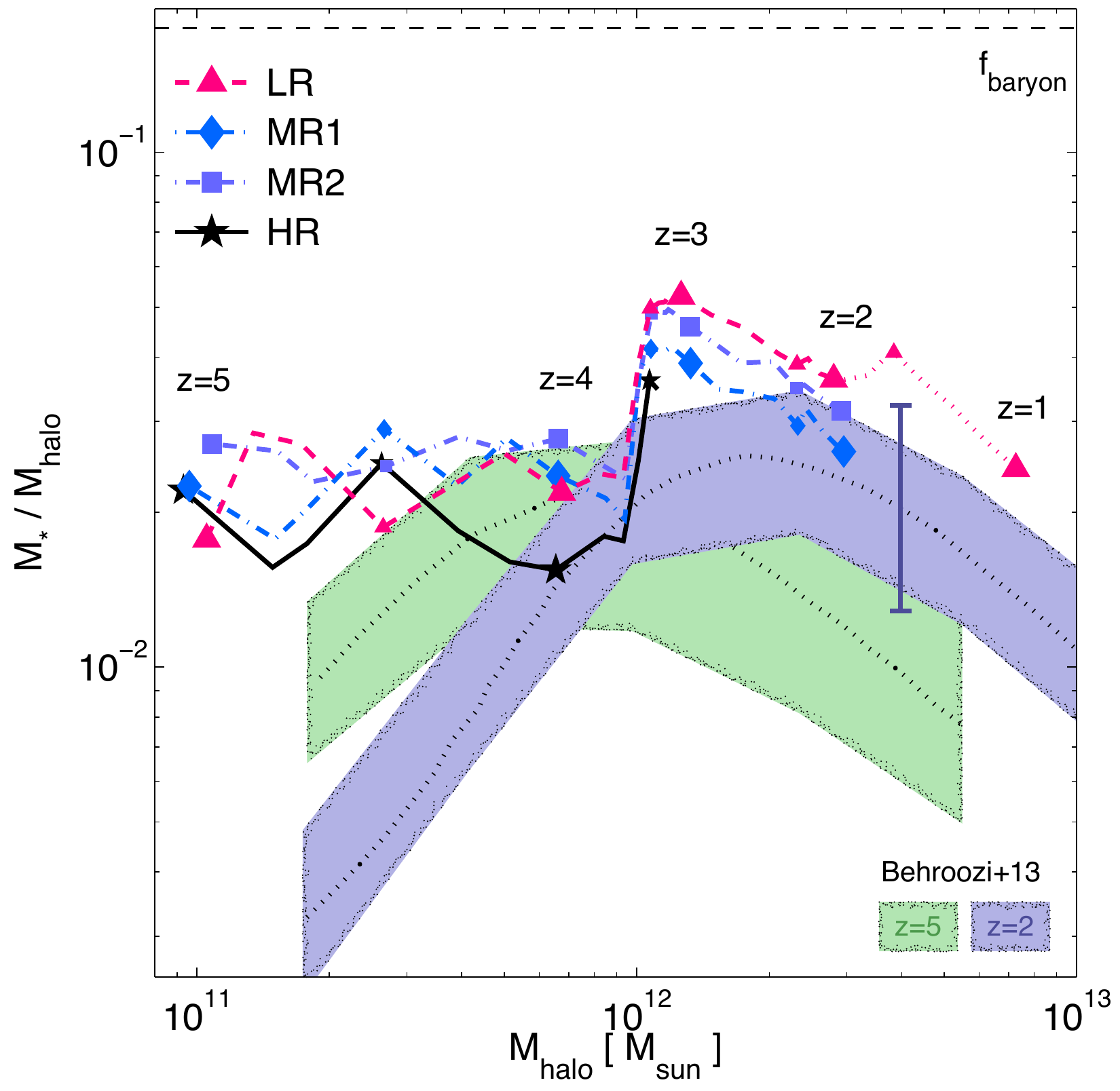}
\end{tabular}
\caption{The stellar-to-virial mass ratio as function of virial mass. Symbols connected with solid, dot-dashed, and dashed lines show the stellar-to-virial mass ratio of the central galaxy in the HR, MR, and LR runs (see legend). The symbols mark the specific redshifts 5, 4, 3, 2, and 1 (from left to right). The thin, horizontal dashed line at the top shows the universal baryon fraction. The dotted lines and shaded regions are the mean value of the stellar fraction and its combined systematic and statistical uncertainty range, respectively, as derived from abundance matching techniques \citep{2013ApJ...770...57B}. The errorbar on the right shows the typical scatter (0.2 dex) for individual halos \citep{2009MNRAS.392..801M, 2013ApJ...771...30R} as comparison. The stellar mass fraction of the simulated galaxy evolves only weakly and is in agreement with abundance matching predictions over the redshift range $z=4$ to $z=2$.}
\label{fig:MsMhalo}
\end{figure}

Hence, we compare in Figure~\ref{fig:MsMhalo} the stellar-to-virial mass ratio of the simulated galaxy in the HR, MR, and LR runs with the relation derived by \cite{2013ApJ...770...57B} using the abundance matching technique \citep{2004MNRAS.353..189V, 2006ApJ...647..201C, 2010MNRAS.404.1111G, 2010ApJ...717..379B, 2010ApJ...710..903M, 2011ApJ...742...16T, 2013ApJ...771...30R, 2013MNRAS.428.3121M}. The HR run of the Argo simulation features the same physics and has the same resolution as the simulation by \cite{2011ApJ...742...76G}, and thus is a solid baseline against we can cross-check our lower resolution runs.

As Figure~\ref{fig:MsMhalo} shows, the stellar-to-virial mass ratio of the simulated central galaxy is in good agreement with the empirically derived estimates by \cite{2013ApJ...770...57B} over the $z=4$ to $z=1$ range. \boldtext{At higher redshifts the stellar-to-virial mass ratio in the simulation lies above abundance matching predictions.} Interestingly, the stellar-to-virial mass ratio is almost constant with a value of $\sim{}2-3\%$, although it shows fluctuations of the order of the scatter in the empirically derived stellar-to-virial mass relation. The rapid increase in the stellar mass at $z\sim{}3.5$ results from a massive starburst and several galaxy mergers that add stellar mass to the central galaxy but do not change the virial mass significantly. Later on, the increase in stellar mass of the galaxy is outpaced by the increase in virial mass and the stellar-to-virial mass ratio drops.

The virial mass among the different runs is in excellent agreement. The stellar mass is also agreeing well, although there is a visible trend of increased stellar mass with decreased numerical resolution, at least after $z=4$. For instance, at $z=3.5$ the stellar mass is $\sim{}40\%$ larger in the LR run than in the HR run. Overall the agreement is reasonable given that gas particle masses change by a factor 64 between the LR and the HR run. We note that our LR run has a significantly better resolution than numerical studies of large ensembles of galaxies in representative cosmological volumes (e.g., \citealt{2013MNRAS.435.2931H, 2013MNRAS.436.3031V, 2013arXiv1312.5462L}). Hence, simulations that use a physical model similar to ours, but at much lower resolution, may run the risk of overestimating the importance of additional feedback mechanisms to reduce stellar masses of galaxies to observed levels. For simulations that reach the resolution of our MR or HR runs a simple supernova-based feedback model \citep{2006MNRAS.373.1074S, 2011ApJ...742...76G} may be all that is needed to reproduce adequate stellar masses for galaxies residing in $\sim{}10^{12.5}$ $M_\odot$ halos at $z=2$. 

Another interesting property of massive quiescent galaxies at redshift $z\sim{}1-3$ is that they are significantly more compact than passive galaxies of similar mass at $z\sim{}0$ (e.g., \citealt{2005ApJ...626..680D, 2006ApJ...650...18T, 2007ApJ...671..285T, 2008ApJ...677L...5V, 2008ApJ...688...48V, 2008ApJ...688..770F, 2008A&A...482...21C, 2010ApJ...709.1018V, 2011ApJ...739L..44D, 2011ApJ...743...96C, 2012ApJ...749..121S, 2013ApJ...773..112C}). This discovery, first based on measuring surface brightness profiles, was subsequently confirmed spectroscopically (e.g. \citealt{2009Natur.460..717V, 2010ApJ...717L.103N, 2011ApJ...736L...9V, 2012ApJ...754....3T, 2012ApJ...755...26O, 2013ApJ...764L...8B, 2013ApJ...771...85V}).

\begin{figure}
\begin{tabular}{c}
\includegraphics[width=80mm]{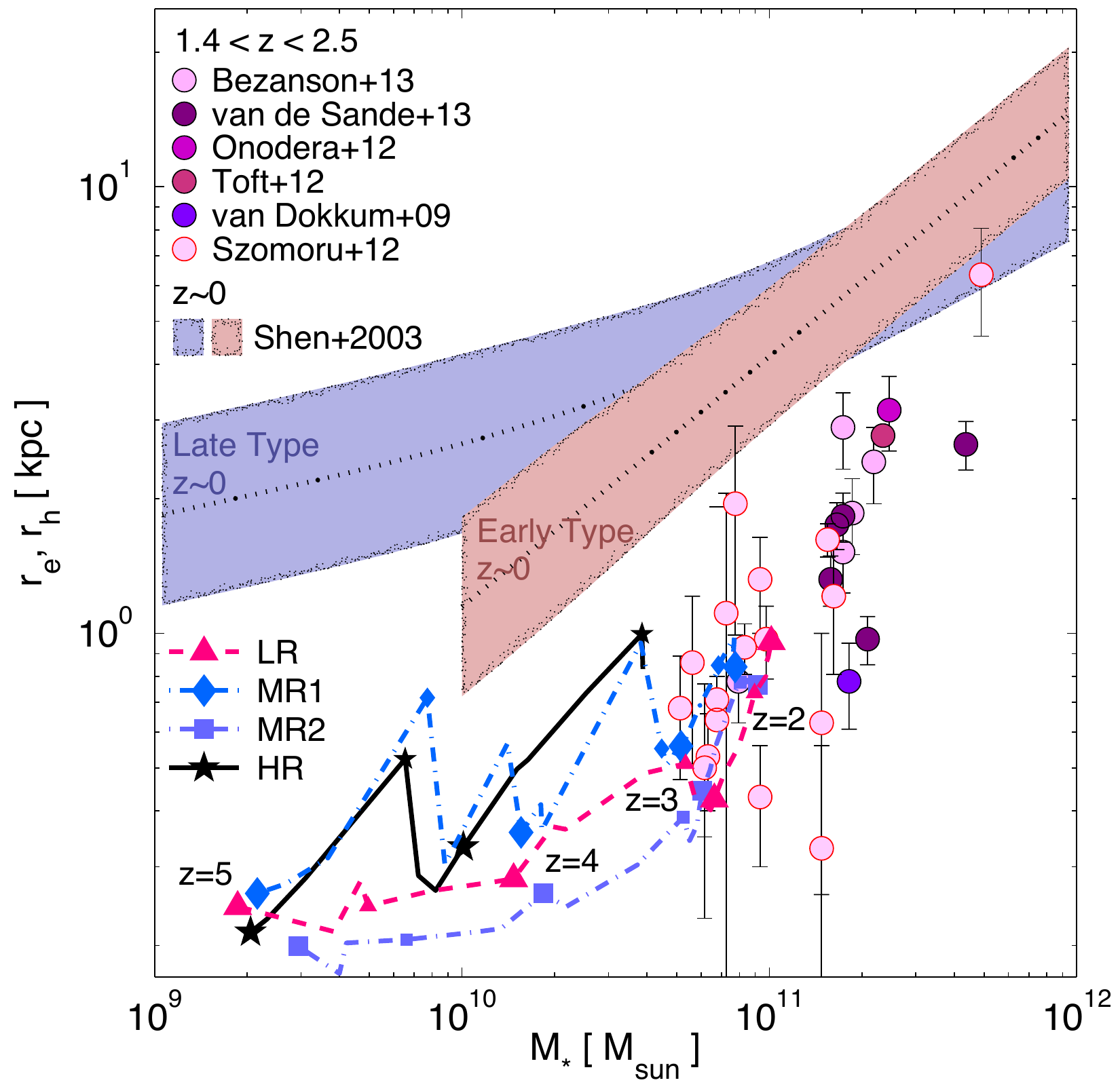}
\end{tabular}
\caption{Stellar mass - size relation of simulated and observed galaxies. Symbols connected with solid, dot-dashed, and dashed lines show the half mass radius of the central galaxy in the HR, MR, and LR runs (see legend). The symbols mark the specific redshifts 5, 4, 3, and 2 (from left to right). Circles with errorbars show the effective radii and stellar masses of $z\sim{}2$ quiescent galaxies \citep{2009Natur.460..717V, 2012ApJ...749..121S, 2012ApJ...754....3T, 2012ApJ...755...26O, 2013ApJ...764L...8B, 2013ApJ...771...85V}. The dotted lines and shaded regions show the observed $z\sim{}0$ mass -- size relation and its scatter \citep{2003MNRAS.343..978S} for galaxies with S\'{e}rsic index $\geq{}$ 2.5 (early types) and $<2.5$ (late types). The size of the simulated galaxy at $z\sim{}2-3$ is in good agreement with observations of quiescent, similarly massive galaxies at $z\sim{}2$. In particular, it is more compact (by 0.3-0.5 dex) than early type galaxies of similar mass in the local Universe.}
\label{fig:MsRh}
\end{figure}

In Figure~\ref{fig:MsRh} we test how well the size of our simulated galaxy fares with those of massive, quiescent galaxies at high $z\sim{}2$. At $z=2$ our simulated galaxy has size of $\sim{}1$ kpc and a stellar mass of $\sim{}10^{11}$ $M_\odot$. This agrees well with the sizes of observed, similarly massive, quiescent galaxies. However, we caution the reader that our sizes are half-mass radii and not ``effective radii''. The latter are typically used in the observational literature and are derived from a parametric fit to the light surface profiles. While keeping these caveats in mind, the similarity of the sizes at $z\sim{}2$ is encouraging. 

\boldtext{At $z>3.5$ the galaxy is already rather compact but (as we show later) its SFR and stellar mass place it on the star forming sequence, i.e., at $z\sim{}4$ it is a compact star forming galaxy. Such galaxies, sometimes called blue nuggets, are possibly an important intermediate stage preceding quenching and the formation of compact, passive galaxies \citep{2011ApJ...742...96W, 2013ApJ...766...15P, 2013ApJ...765..104B}.}

The size of our galaxy at $z\sim{}2-3$ changes little among the runs, again showing that we have reasonably well converged to a final answer by that time. At higher redshifts, however, there are noticeable differences of up to a factor of 3 between the runs. \boldtext{Note, that the half mass radius is barely resolved at $z>4$ even in our HR run.} Overall we find a tendency for larger size with increasing resolution. The trend is possibly caused by the enhanced efficiency of supernova driven winds to remove low angular momentum material with increased numerical resolution, as seen in previous work (e.g., \citealt{2010Natur.463..203G, 2011ApJ...742...76G}). \boldtext{In addition, a higher numerical resolution results in a larger number of resolved satellite galaxies and, hence, may drive larger sizes via an increase in the minor merging rate.}

The size evolution in our highest resolution run (HR) and in the MR1 medium resolution run shows some interesting trends. First, the half-mass radius grows between $z=5$ and $z=4.5$ as a consequence of star formation in the outer region of the galaxy. Then at $z=4.5$ the galaxy undergoes a significant stellar merger and the half-mass radius drops significantly. This behavior repeats itself, when the galaxy size increases between $z\sim{}4.3$ and $z=3.4$, only to start to drop at $z\sim{}3.5$, when (as we show later) the galaxy undergoes a critical event in its history. Between $z=3.5$ and $z=3$ the evolution in half-mass radius and stellar mass is minimal. After $z=3$ the galaxy growth quickly in size, but add little stellar mass. If we trust the size evolution in the HR and MR1 runs at $z\geq{}3.5$, then this points to a picture in which galaxies grow in size while being on the star forming sequence (via star formation), shrink when engaging in wet mergers, and then finally grow via gas-poor mergers after leaving the star forming sequence. 
 
So far we have tested how the stellar mass and the structural properties of the simulated massive galaxy compare with observational data. Another crucial check is to test the properties of the hot gas surrounding the galaxy. The entropy profile is of particular importance, because, it measures the cooling time of the circum-galactic gas.

\begin{figure}
\begin{tabular}{c}
\includegraphics[width=80mm]{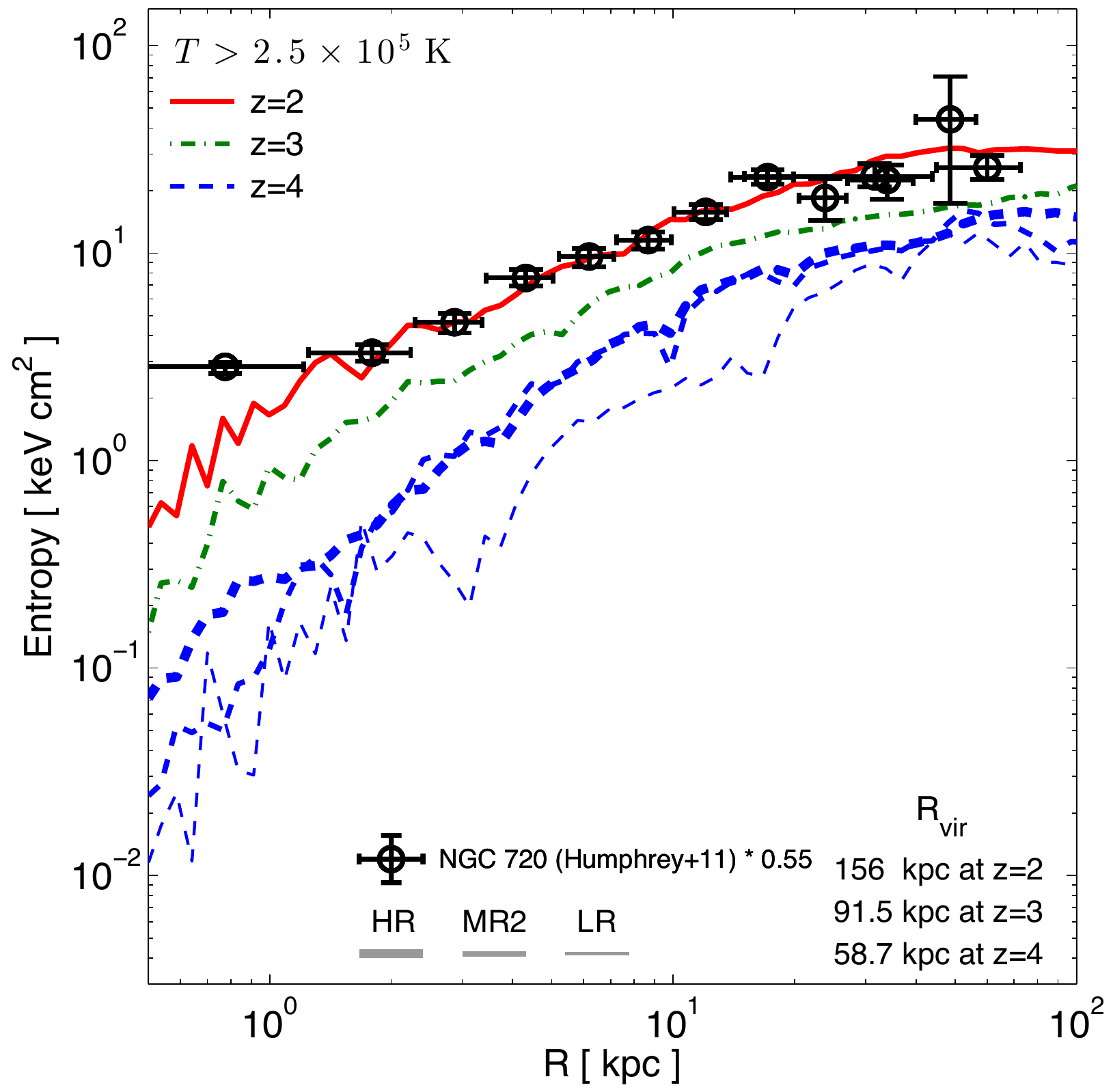} 
\end{tabular}
\caption{Entropy profile ($S=T/n_{\rm e}^{2/3}$, where $n_{\rm e}$ is the number density of electrons) radially averaged in spherical shells. Red solid, green dot-dashed and blue dashed lines show the entropy profile of hot gas ($T>2.5\times{}10^5$ K) around the simulated massive galaxy at redshift 2, 3, and 4, respectively. At $z=4$ the thickness of the line corresponds to the resolution of the simulation, see legend. The lines at $z=2$ and $z=3$ are from the MR2 run. Circles with error bars show the entropy profile (rescaled to 55\% to match the normalization) of the hot gas around the nearby quiescent elliptical galaxy NGC 720 \citep{2011ApJ...729...53H}. The shape of the entropy profiles of the simulated galaxy at $z=2$ and the observed galaxy at $z\sim{}0$ agree well, except maybe in the central kpc of the galaxy. The entropy is below the critical value of $\sim{}100$ keV cm$^2$ \citep{2001Natur.414..425V, 2004ApJ...608...62S} indicating that the gas cooling time is shorter than the Hubble time. Supernova feedback as well as gravitational heating balance the cooling rate of the hot gas halo and avoid excessive net cooling rates at $z\sim{}2$. At low redshifts additional feedback sources, e.g., a radio mode AGN, are likely required to maintain the low SFR of the central galaxy.}
\label{fig:Entropy}
\end{figure}

In Figure~\ref{fig:Entropy} we show the entropy profile at various redshifts and compare it with the entropy profile of the nearby massive, quiescent elliptical galaxy NGC 720 \citep{2011ApJ...729...53H}. NGC 720 has a stellar mass of $\sim{}10^{11}$ $M_\odot$ and is therefore a potential low redshift counterpart of our simulated galaxy at $z=2$. As the figure demonstrates, the entropy profile of the simulated galaxy at $z=2$ and the entropy profile of NGC 720 have the same shape and almost the same normalization (to within a factor of 2). The intriguing property of NGC 720 is that it somehow prevents the overcooling problem without increasing the entropy of the gas to the critical value of $\sim{}100$ KeV cm$^2$, above which cooling would take longer than a Hubble time. Hence, it is a counter example to the picture in which star formation is quenched because of a previous quasar mode heating phase (e.g., \citealt{2004ApJ...608...62S}).

In fact, the cooling times of the gas are quite short, especially at small distances from the center of the galaxy. At $z=2$, the cooling time is only $\sim{}10$ Myr for $r=1$ kpc, while it reaches several Gyr ($\sim{}$a Hubble time) at $r=100$ kpc. This indicates that to avoid overcooling, gravitational heating as well as heating from supernovae have to replenish a large fraction of the cooling losses. We note that the cooling time generally exceeds the dynamical time by a factor of a few at each radius. The heating mechanisms implemented in the Argo simulation might not suffice in preventing overcooling at late times. In the real Universe, additional heating sources, e.g., an AGN in radio mode, may step in and supplant the decreasing contribution from gravitational heating (fewer mergers) and supernova heating (fewer core collapse supernovae).

The entropy profile of the MR and HR runs are converged except maybe in the central kpc. This difference, however, may explain the somewhat larger stellar mass in the MR2 run compared with the HR run as most of the star formation takes place in the central kpc. We also note that the entropy is lower in the LR run (by a factor of $\sim{}2-3$ at most radii) although it matches the entropy of the MR2 run in the central region. We interpret the smaller entropy in the LR and MR2 runs as a consequence of the reduced efficiency of supernova feedback at lower resolution and with a lower star formation threshold (e.g., \citealt{2011ApJ...742...76G, 2013ApJ...766...56M}).

\begin{figure*}
\begin{tabular}{cc}
\includegraphics[width=80mm]{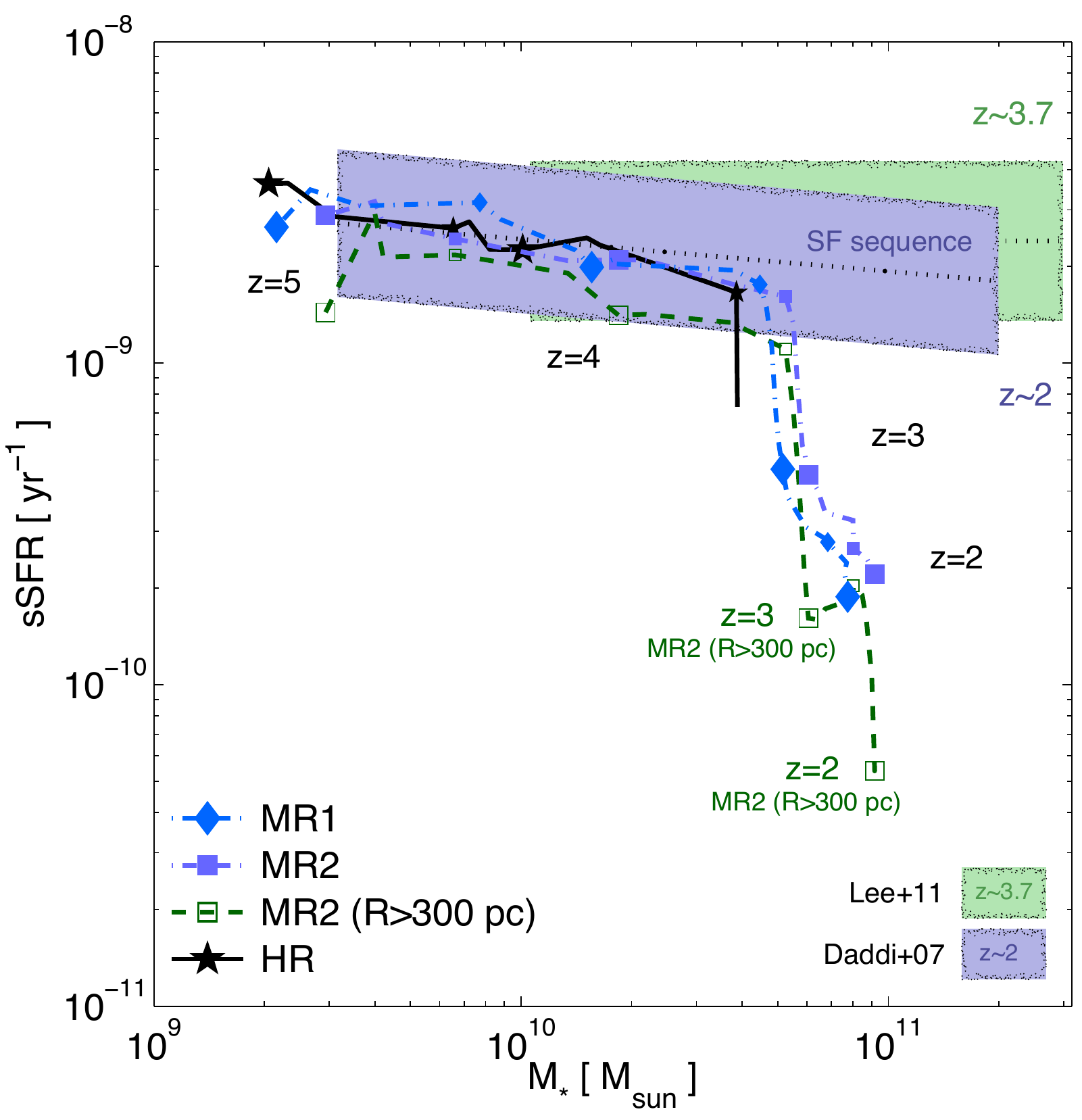} &
\includegraphics[width=80mm]{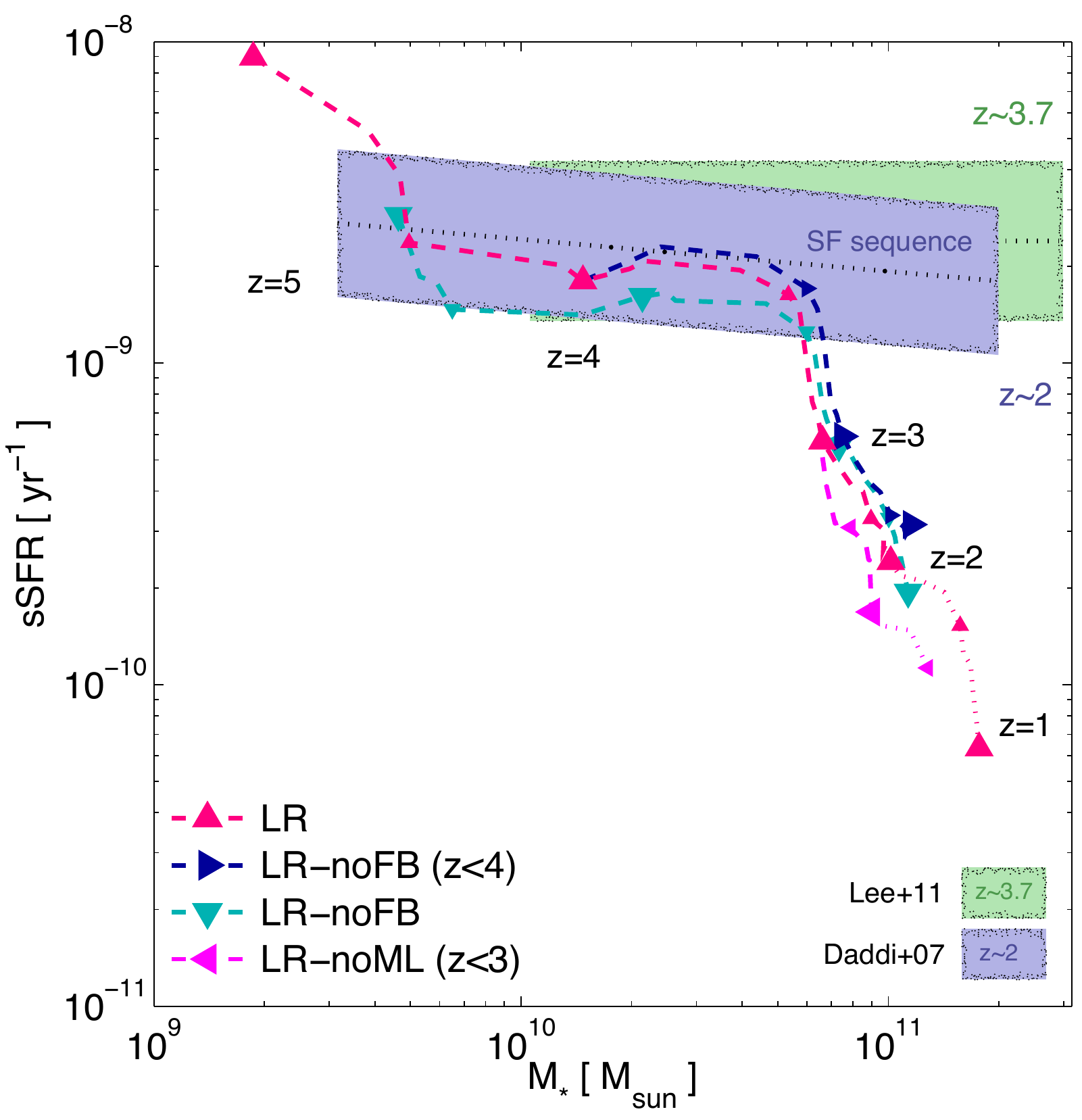} \\
\end{tabular}
\caption{Evolution of the sSFR. (Left panel) Evolution of the sSFR of the central galaxy in the HR (black solid line, star symbols) and the MR (dot-dashed lines, square and diamond symbols) runs. The line connecting empty squares show the sSFR of the central galaxy when the star formation and stellar mass within the central 300 pc are excluded. (Right panel) Evolution of the sSFR of the central galaxy in the LR run (red lines, upward-pointing triangle symbols) and in various re-simulations of the LR run. The blue line with right-pointing triangles is a run that starts with the $z=4$ snapshot of the LR run and the continues down to $z=2$, but with SN feedback turned off. Similarly, the green line with downward-pointing triangles is a re-simulation of the LR run but without SN feedback from the start. The magenta line with left-ward pointing triangles is a re-simulation that starts with the $z=3$ snapshot of the LR run but with stellar mass loss turned off. In both panels large symbols mark redshifts 5, 4, 3, 2 and 1 (from left to right), small symbols mark redshifts 4.5, 3.5, 2.5, 1.5. The SFRs are measured within a 20 Myr time interval and a 5-point moving average is applied to the sSFR to reduce short term fluctuations and highlight the overall evolutionary trend. Dotted lines and shaded regions show the mean location and the 1-$\sigma$ scatter, respectively, of the observed star formation sequence at $z\sim{}3.7$ \protect\citep{2011ApJ...733...99L} and $z\sim{}2$ \protect\citep{2007ApJ...670..156D}. Stellar masses are converted to a \protect\cite{2003PASP..115..763C} IMF.  The central galaxy evolves along the observed star forming sequence until $z\sim{}3.5$. At $z\sim{}3.5$ the sSFR drops by a large factor ($\sim{}5-10$) within a few 100 Myr, resulting in a massive galaxy with strongly suppressed star formation activity.}
\label{fig:MssSFR}
\end{figure*}

\section{On Leaving the Star Forming Sequence}
\label{sect:LeavingSFS}

In the previous section we traced the simulated massive galaxy across $z=5-2$ and analyzed how its properties change over this redshift range, finding significant evolution. We showed that stellar mass, stellar-to-virial mass ratio, size, and colors of the simulated galaxy are in good agreement will those of massive ($M_*\sim{}10^{11}$ $M_\odot$) quiescent galaxies observed at high redshift. We will now discuss the evolution of the SFR and the sSFR of the simulated galaxy.

We show in Figure~\ref{fig:MssSFR} the sSFR of the simulated galaxy as function of its stellar mass. At $z\geq{}3.5$ the galaxy lies on the observed star forming sequence (sometimes called the ``main sequence'') with a sSFR of $\sim{}2\times{}10^{-9}$ yr$^{-1}$. We should point out that this is an important achievement of the Argo simulation. Indeed, published galaxy simulation studies, especially those with resolutions worse than our LR run, often obtain sSFRs that lie off the observed relation by a factor 2-3 (e.g., \citealt{2011MNRAS.410.1703F, 2013MNRAS.435.2931H}).

At $z=3.5$ the galaxy leaves, rather abruptly, the star forming sequence. The sSFR drops by a large factor ($\sim{}5-10$) within a few hundred Myr.  At $z=2$, the sSFR is only $\sim{}10^{-10}$ yr$^{-1}$, i.e., more than an order of magnitude lower than the sSFRs of galaxies on the star forming sequence. The abrupt suppression of the sSFR at $z\sim{}3.5-3$ is reproduced in all our runs. The drop is somewhat steeper in the MR and HR runs compared with the LR run. We interpret this result as a consequence of supernova feedback being somewhat more efficient in our higher resolution runs, see also section \ref{sect:CompObs}. 

However, \emph{the drop of the sSFR is not triggered by supernova feedback}. This is shown in the right panel of Figure~\ref{fig:MssSFR} where we test how the sSFR evolves after turning off the feedback at $z=4$. We find little difference whether supernova feedback is active or not. 

Feedback nonetheless plays two important roles. First, feedback at early times ($z\gg{}4$) is required to prevent overcooling and to reproduce sSFRs consistent with observations. We illustrate this point in Figure~\ref{fig:MssSFR} where we show re-simulations of the LR run without feedback. The run in which we switch off supernova feedback after $z=4$ differs little from the default LR run as noted above. \boldtext{In contrast, the run without any supernova feedback (LR-noFB) results in too high a stellar mass which outweighs the also somewhat higher SFR, especially at early times. As a consequence, the sSFR at $z>3.5$ is significantly reduced in this run.}

Figure~\ref{fig:MssSFR} shows that the sSFR levels off at $\sim{}10^{-10}$ yr$^{-1}$ (at $z=2$) and only slowly decreases as time proceeds. As we will demonstrate in section \ref{section:OriginQuenching}, this floor in the sSFR at high redshift is sustained by the accretion of cool gas from large distances. However, one may wonder whether mass loss from evolved stars (included in our models) contributes at a significant level (e.g., \citealt{2013arXiv1308.4132F}). The right panel of Figure~\ref{fig:MssSFR} shows a re-simulation of the LR run, but with the stellar mass loss switched off after $z=3$. In this case the sSFR levels off at a somewhat smaller value compared with the default LR run. However, the difference is small (smaller than a factor 2) proving that the $\sim{}10^{-10}$ yr$^{-1}$ floor of the sSFR at $z=2$ is not related to a replenishment of the ISM by mass loss from an evolved stellar population.

\begin{figure*}
\begin{tabular}{cc}
\includegraphics[width=80mm]{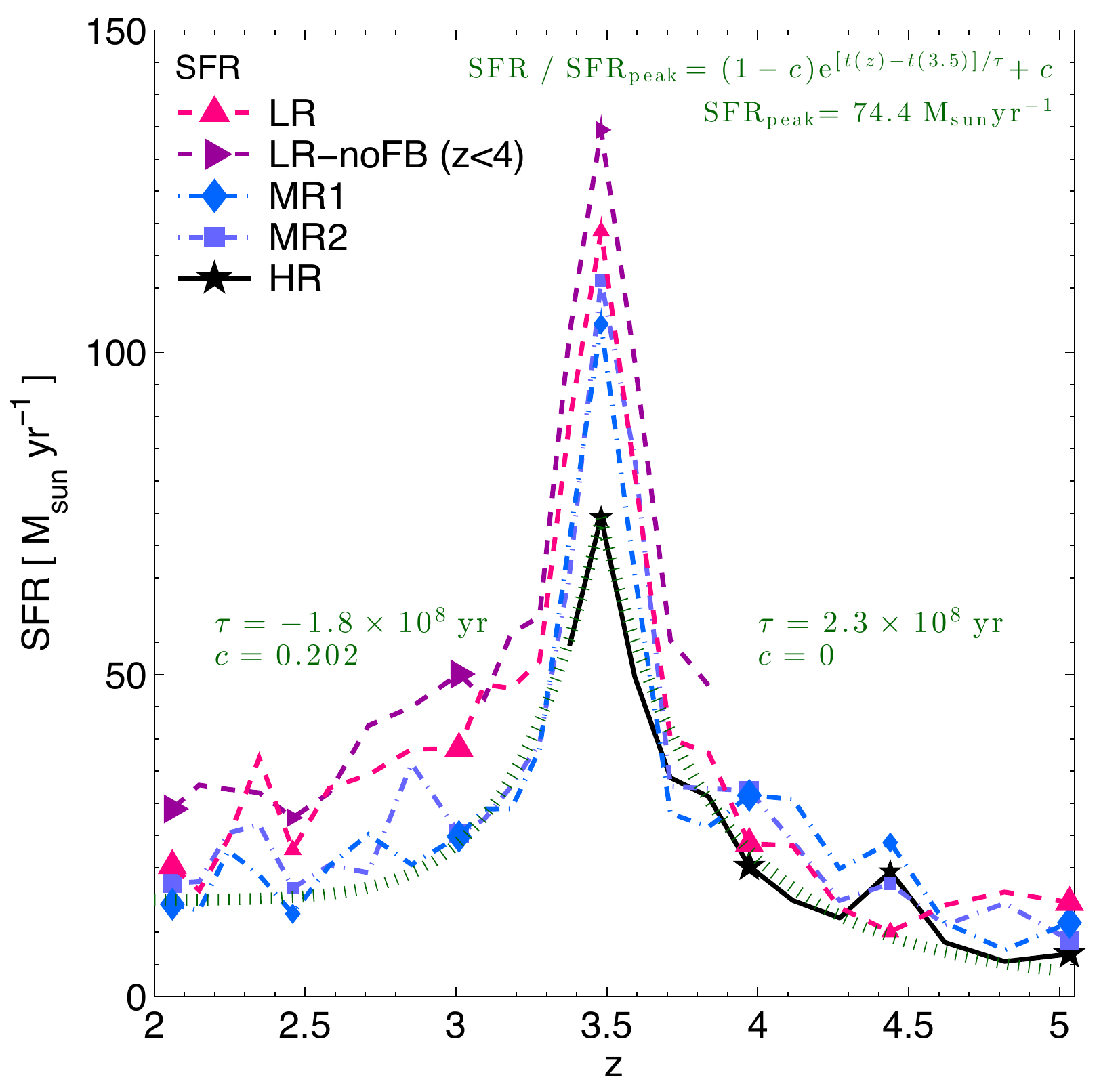} &
\includegraphics[width=78.8mm]{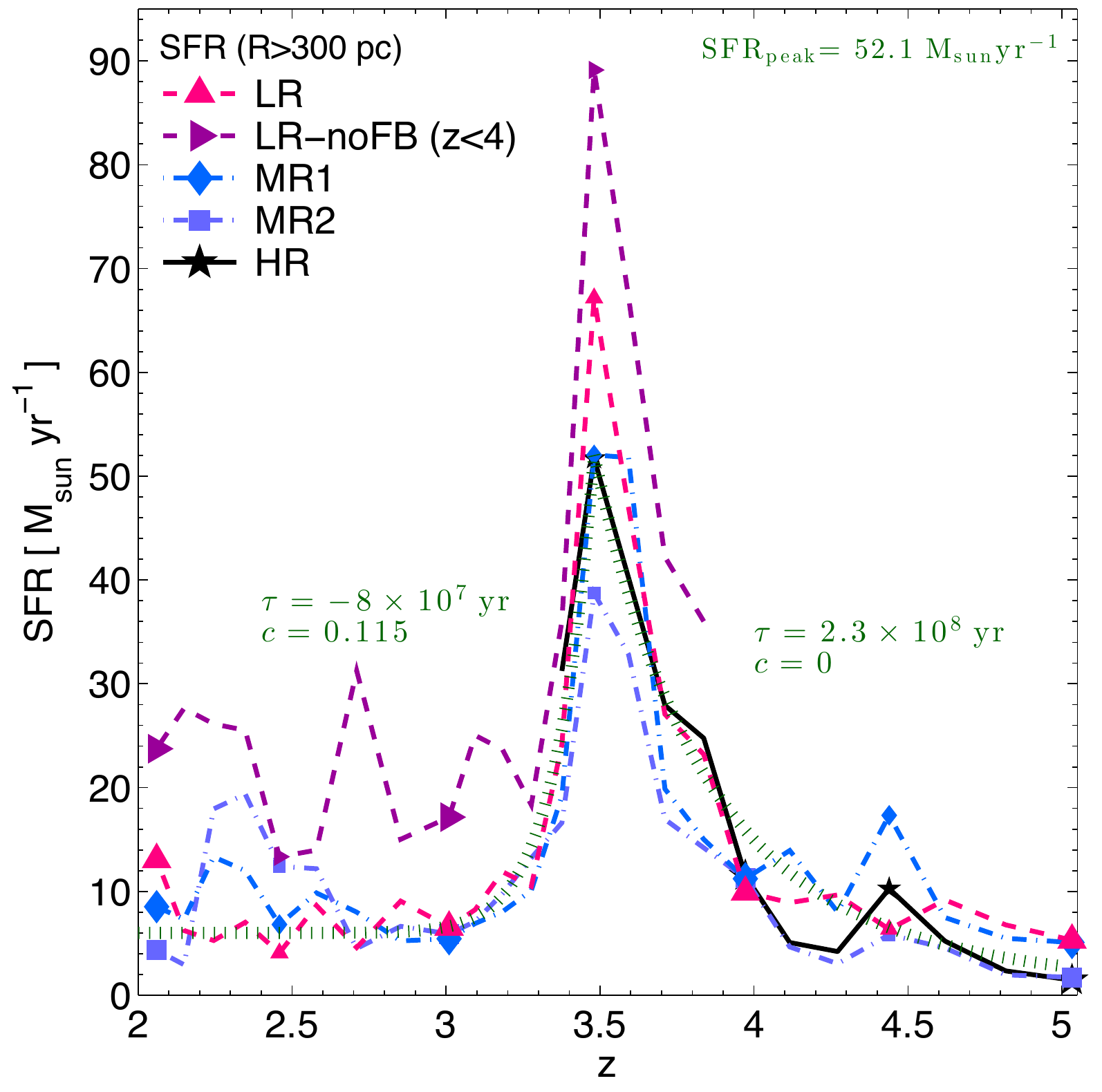}
\end{tabular}
\caption{Star formation history of the simulated massive galaxy. (Left panel) SFR within a radius of 12 kpc excluding contributions from satellite galaxies. (Right panel) SFR within a shell with a 0.3 kpc inner radius and a 12 kpc outer radius. Symbols connected with solid, dot-dashed, and dashed lines show the SFR of the central galaxy in the HR, MR, and LR runs (see legend). The SFR increases exponentially until $z=3.5$ after which it declines. At $z\sim{}2-3$ a non-negligible fraction of the star formation takes place within the central 0.3 kpc.}
\label{fig:SFhist}
\end{figure*}

The left panel of Figure~\ref{fig:SFhist} shows the star formation history of the massive galaxy traced through time. The SFR increases steeply from $\sim{}10$ $M_\odot$ yr$^{-1}$ at $z=5$ to $\sim{}100$ $M_\odot$ yr$^{-1}$ at $z=3.5$. Afterwards the SFR decreases and settles at $\sim{}15$ $M_\odot$ yr$^{-1}$ at $z=2$ (for the MR runs). 

Comparing the LR, MR and HR runs, we find differences in the SFR (close to a factor 2), especially around the peak of SFR at $z\sim{}3.5$ and in the subsequent star formation activity at $z\sim{}2 - 3.5$. The general trend is that the SFR is reduced in the runs with higher spatial resolution, i.e., in the runs with more efficient stellar feedback. However, a significant fraction of the star formation takes place in the central softening length of the simulation and, hence, is unresolved. We thus show in the right panel of Figure~\ref{fig:SFhist} the SFR that occurs outside the central 300 pc. In this case the predictions of the various runs are well converged.

The SFR can be approximated by an exponential with constant offset, i.e.,
\begin{equation}
{\rm SFR}(t) = {\rm SFR}(t_*)\left[ (1-c) e^{\frac{t-t_*}{\tau}} + c\right].
\label{eq:SFRexponential}
\end{equation}
Here, $t_*=t(z=3.5)=1.9$ Gyr, $c$ is an offset, and $\tau$ is the star formation time scale. The value of $\tau$ is positive for an exponentially growing SFR and negative for an exponentially declining SFR. The absolute value of $\tau$ measures the gas accretion time ($M_{\rm gas}/\dot{M}_{\rm gas,acc}$) and the gas depletion time ($M_{\rm gas}/{\rm SFR}$) in the  growing and declining phase, respectively, see appendix \ref{sect:FuncFormSFH}.

At $z>3.5$ the SFR increases roughly exponentially with time ($c=0$). The e-fold time is $\tau\sim{}2.3\times{}10^8$ yr for the HR run. This e-fold time corresponds to a sSFR of $1/\tau \sim{} 4\times{}10^{-9}$ yr$^{-1}$, approximately a factor 2 larger than the actual sSFR of the galaxy, see Figure~\ref{fig:MssSFR}. However, the $1/\tau$ estimate ignores the contribution of stellar mass growth via mergers\footnote{It also ignores stellar mass loss, which, when properly taken into account, increases the discrepancy by a few tens of a percent.}. Indeed, the simulated galaxy experiences a number of significant merger events between $z=5$ and $z=3.5$. It undergoes a major galaxy merger at $z=4.4$ (the stellar mass ratio is $1:2.5$). It is involved further in a $1:10$ merger at $z=3.9$, a $1:7$ merger at $z=3.7$, and a $1:5$ merger at $z=3.5$. The overall stellar mass growth provided by galaxy mergers is $\sim{}1.7-2\times{}10^{10}$ $M_\odot$, comparable to the mass growth by in situ star formation ($\sim{}1.6\times{}10^{10}$ $M_\odot$). The mass acquisition via mergers explains why the sSFR is actually a factor two smaller than the estimate $1/\tau$.

The exponential growth of the SFR at $z>3.5$ also ensures that the sSFR remains roughly constant from $z=5$ to $z=3.5$. Naively, one might expect that a peak in the SFR translates into a peak in the sSFR, i.e., in an upward deviation from the star forming sequence at $z\sim{}3.5-4$. However, since the exponential function is proportional to its own derivative the sSFR is constant for a galaxy that grows most of its mass via an exponentially increasing SFR. 

\begin{figure*}
\begin{tabular}{cc}
\includegraphics[width=80mm]{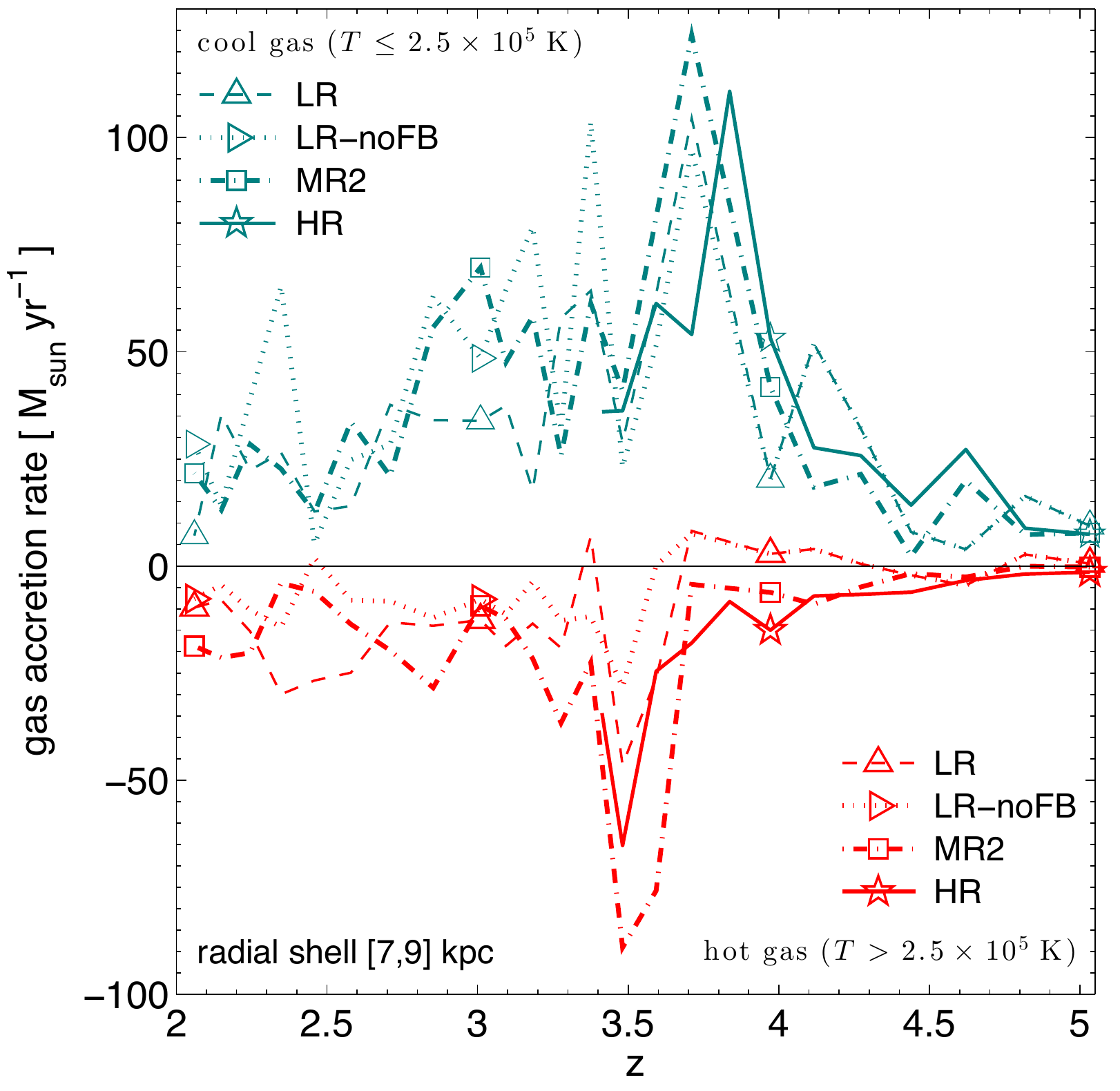} & 
\includegraphics[width=79.2mm]{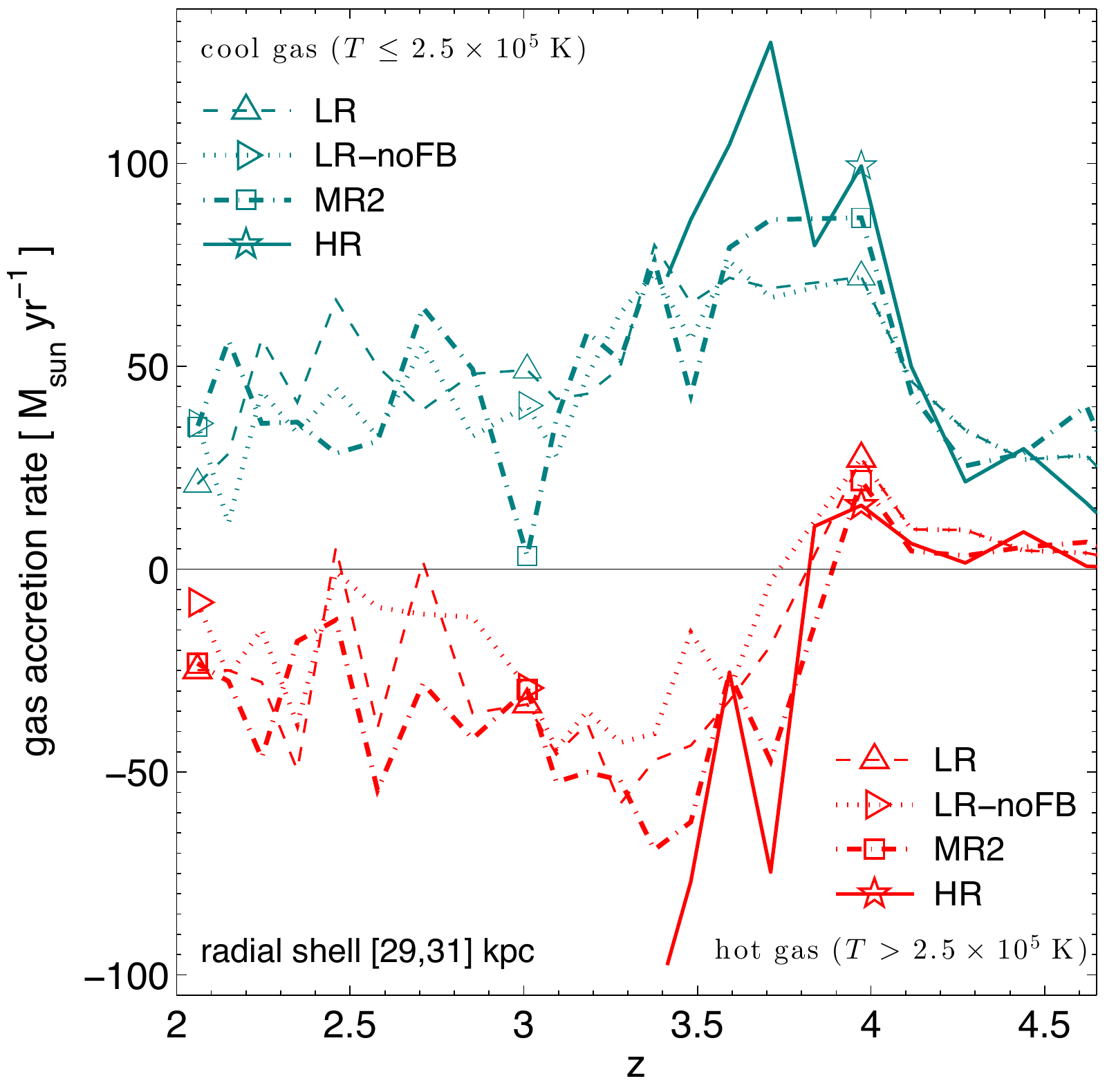}
\end{tabular}
\caption{Accretion rates of cool ($T\leq{}2.5\times{}10^5$ K) and hot ($T>2.5\times{}10^5$ K) gas as measured in spherical shells centered at proper 8 kpc (left panel) and 30 kpc (right panel) for the LR, MR, and HR runs (see legend). Cool gas flows toward the central galaxy, while hot gas, especially at $z\lesssim{}4$, preferentially flows outward. At $z\sim{}2-3$ the outflow rate of hot gas balances a large fraction of the inflow rate of the cool gas at each radius, resulting in little net accretion of gas toward the galaxy.}
\label{fig:accretion rates}
\end{figure*}

The star formation activity peaks at $z=3.5$. In the HR run, the SFR reaches $\sim{}75$ $M_\odot$ yr$^{-1}$ ($\sim{}52$ $M_\odot$ yr$^{-1}$ if we exclude the central 300 pc). Afterwards the SFR declines rapidly (approximately as an exponential with constant offset). The SFR has an e-fold time of $\sim{}200$ Myr (100 Myr if we exclude non-central star formation). Hence, within a few 100 Myr the SFR decreases to the much lower levels of $\sim{}15-20$ $M_\odot$ yr$^{-1}$ ($\sim{}5-10$ $M_\odot$ yr$^{-1}$ if we exclude non-central star formation).

Although the galaxy leaves the star forming sequence at $z\sim{}3.5$ (Figure~\ref{fig:MssSFR}), and reduces its SFR within a few 100 Myr (Figure~\ref{fig:SFhist}), it would not be identified as a quiescent galaxy based on the UVJ classification (Figure~\ref{fig:UVJ}) until $z\sim{}2$. We suggest that this classification be improved by extending the region of quiescent galaxies in the UVJ diagram to the lower left.

\section{The origin of star formation quenching}
\label{section:OriginQuenching}

The sSFR of the simulated massive galaxy decreases steeply at $z\sim{}3.5$. We showed in the previous section that this decline is not triggered by feedback processes. Instead, we will now demonstrate that the reduction of the gas supply from the cosmic web, a process we dub ``cosmological starvation'' (or ``cosmic starvation''), is to blame. 

In Figure~\ref{fig:accretion rates} we plot the accretion rates of cool ($T\leq{}2.5\times{}10^5$ K) and hot ($T>2.5\times{}10^5$ K) gas through spherical shells centered on the galaxy. The cool gas accretion rate clearly mimics the star formation history, both in shape and normalization, especially if the accretion rate is measured close to the galaxy (here at 8 kpc distance). The delay ($\sim{}100$ Myr) between the peak in the cool gas accretion rate ($z\sim{}3.7-3.8$) and the peak in the SFR ($z\sim{}3.5$) indicates the time it takes the cool gas to reach the galaxy and participate in star formation. 

Hot gas flows towards to galaxy only at $z>4$ and only if it is located sufficiently far from the massive galaxy. At $z<4$ and/or at small radii, the bulk motion of hot gas is to move away from the galaxy. The outflow rate at small radii tracks the SFR rather precisely with a mass loading factor of the order of unity. The outflow rate peaks at $z\sim{}3.5$ with values of $50-100$ $M_\odot$ yr$^{-1}$. At larger radii the outflow rate evolves more gradual with time. At late times the outflow rate of the hot gas balances a substantial fraction of the cool gas accretion rate. Hence, the net gas accretion at late times is rather small (typically no more than 15 $M_\odot$ yr$^{-1}$).

The cool gas accretion rates agree well among the various runs. The outflow rates of the hot gas differ noticeably, however. In particular, the LR run shows weaker hot gas outflows compared with the MR and HR runs. Again, this is consistent with our overall finding that the supernova feedback scheme we use in the Argo simulation is less effective at low resolution. Nonetheless, there is a bulk radial motion of the hot gas even if we turn off the feedback completely. We suspect that some fraction of the hot gas bulk motion is related to the push of the thermal pressure of hot gas and to sloshing motions of the gas induced by infalling substructures. Supernova feedback, however, is responsible for the much stronger hot gas outflows in the MR and HR runs. We note that the hot gas outflows do not appear to strongly affect the star formation activity of the massive galaxy at $z=3.3-5$, when much of the stellar mass of the massive galaxy is built. At late times ($z=2-3.3$) the larger outflow rates in the MR2 run compared with the LR run result in a reduced star formation activity, but the overall impact on the $z=2$ stellar mass is relatively small, see Table~\ref{tab:Results}.

\begin{figure*}
\begin{tabular}{cc}
\includegraphics[width=80mm]{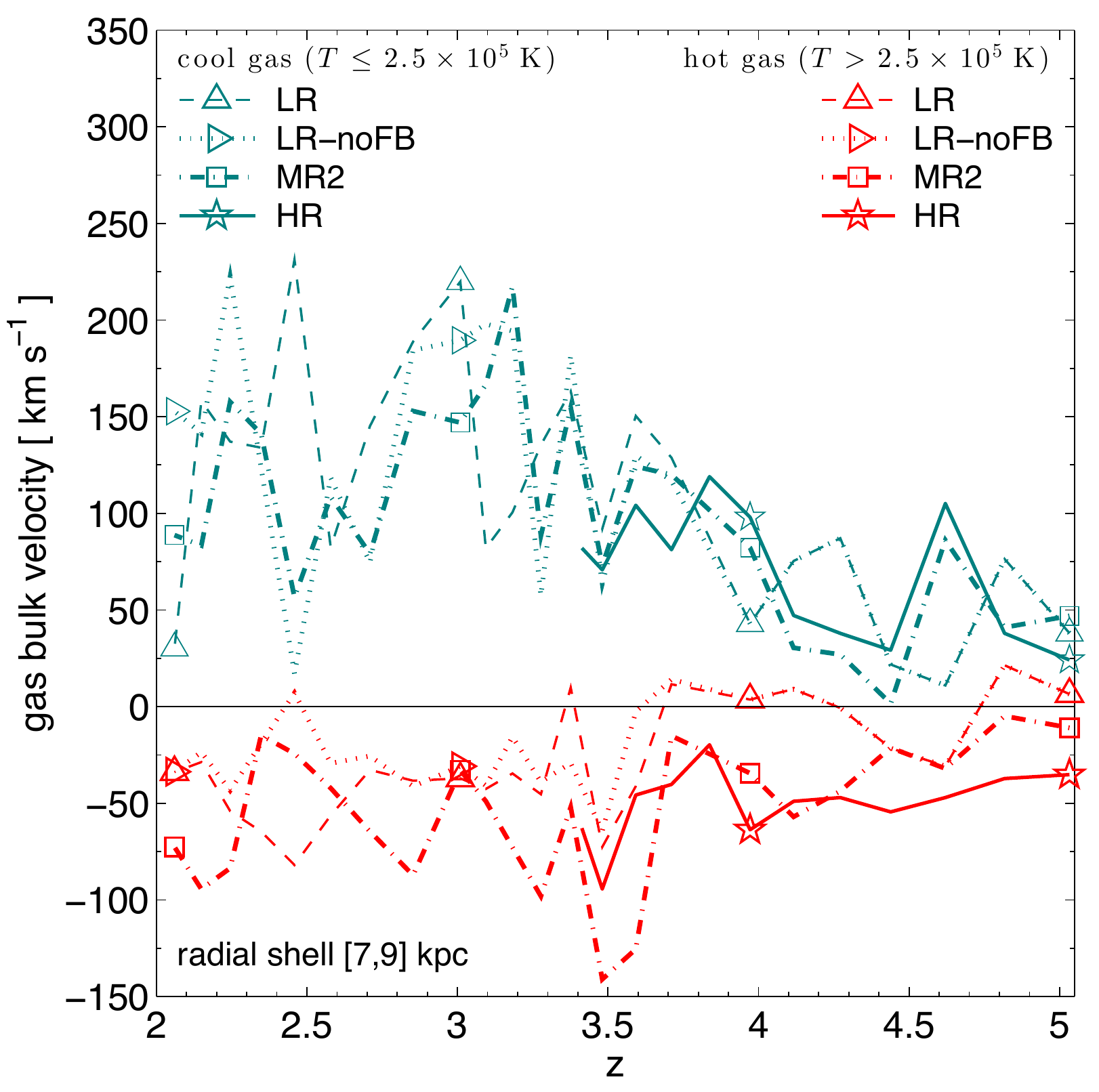} & 
\includegraphics[width=79.2mm]{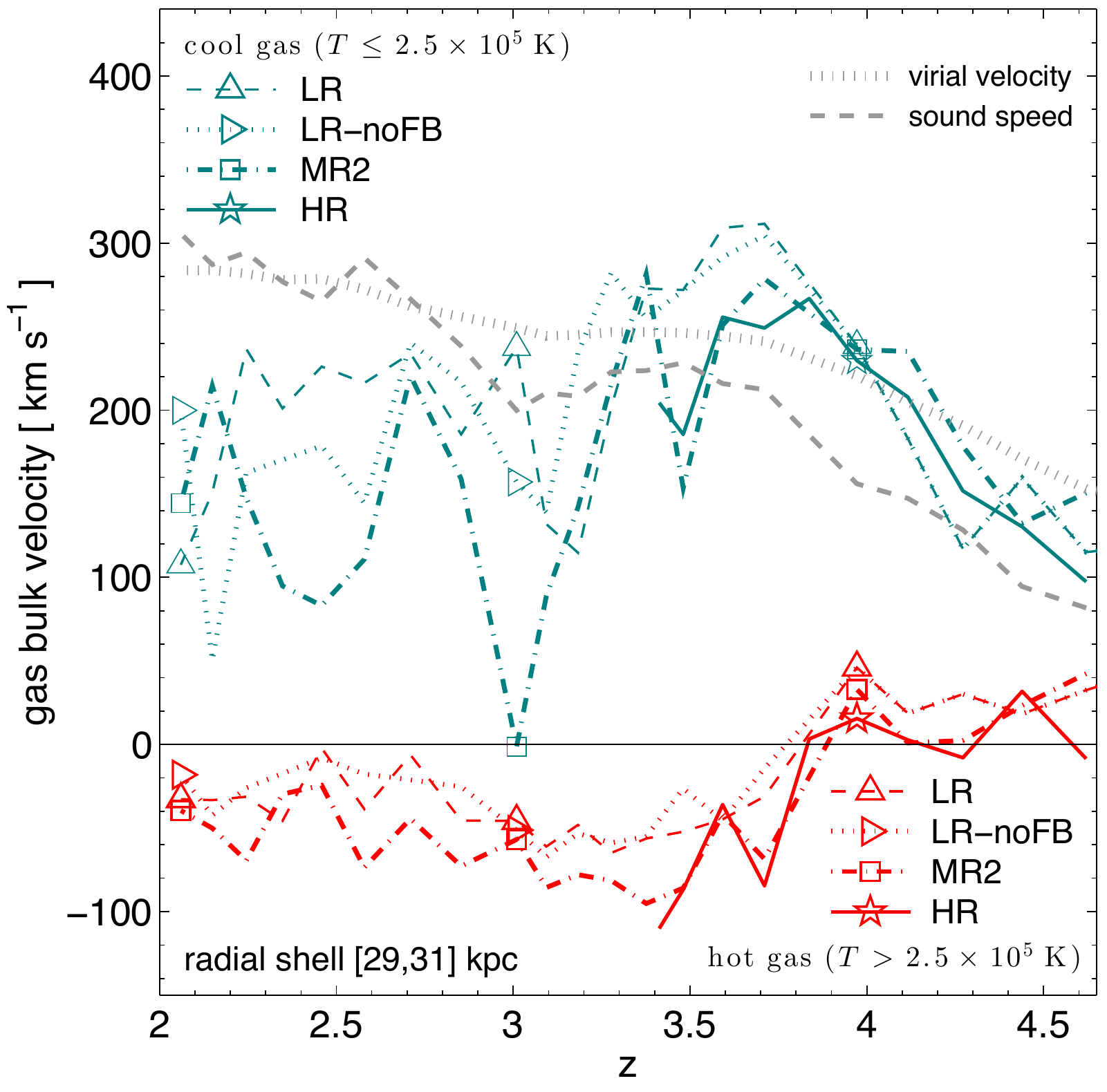}
\end{tabular}
\caption{Bulk radial motion of cool ($T\leq{}2.5\times{}10^5$ K) and hot ($T>2.5\times{}10^5$ K) gas measured in spherical shells centered at proper 8 kpc (left panel) and 30 kpc (right panel) for the LR, MR, and HR runs (see legend). Thick dashed lines show the virial velocity ($[GM_{\rm halo}/R_{\rm halo}]^{1/2}$; upper line) and the adiabatic sound speed ($[\frac{5}{3}k_{\rm B} \overline{T} / \mu]^{1/2}$, with $\mu=0.6 m_{\rm H}$ and $\overline{T}$ the mass-weighted average temperature in the shell; lower line), respectively. Bulk velocities are measured in 2 kpc thick spherical shells. At $z>3.5$ cool gas at large distances flows inward supersonically w.r.t. the hot halo surrounding the galaxy. Near the center of the halo and at late times the inflow rate is sub-sonic. At $z<3.8$ and/or at small radii hot gas is moving radially outward with an average bulk velocity of $\sim{}50-100$ km s$^{-1}$ (in the MR and HR runs).}
\label{fig:accretion velocities}
\end{figure*}

To understand the nature of the cool gas inflows and the hot gas outflows we plot in Figure~\ref{fig:accretion velocities} their bulk velocities. The radial velocity of the cool gas increases until $z\sim{}3.5-3.8$. By $z\sim{}3.5$ the inflow velocity is $\sim{}250-300$ km s$^{-1}$ at a 30 kpc distance and $\sim{}100-150$ km s$^{-1}$ at a 8 kpc distance from the center of the halo, respectively.
 The bulk flow of the hot gas is directed away from the massive galaxy, with flow velocities of $\sim{}50-100$ km s$^{-1}$. These velocities are relatively small indicating that supernova feedback in our simulation probably primarily heats the hot gas halo and induces a pressure supported expansion of the hot halo atmosphere surrounding the central galaxy. 
 
The inflow velocity of the cool gas stalls at late times. Combined with the sharp decrease in the net mass accretion rate this indicates that the mean density of the gas accreted at late times is lower. A lower density allows the virial shock to grow to larger radii (\citealt{2003MNRAS.345..349B}, equation 29). The virial shock heats a much larger volume of gas surrounding the galaxy to the virial temperature of a few $10^6$ K, significantly reducing the cooling rate of gas at large distances from the galaxy. The virial shock thus helps in reducing the inflow rate of cool gas from large to small radii. However, ultimately these events are set in place by the reduction of the gas accretion rates at large radii and not by the formation of the virial shock. In fact, a virial shock forms well before $z=4$ when the simulated galaxy is forming stars at an increasing rate and its sSFR matches those of observed galaxies on the star forming sequence.

\begin{figure*}
\begin{tabular}{cc}
\includegraphics[width=80mm]{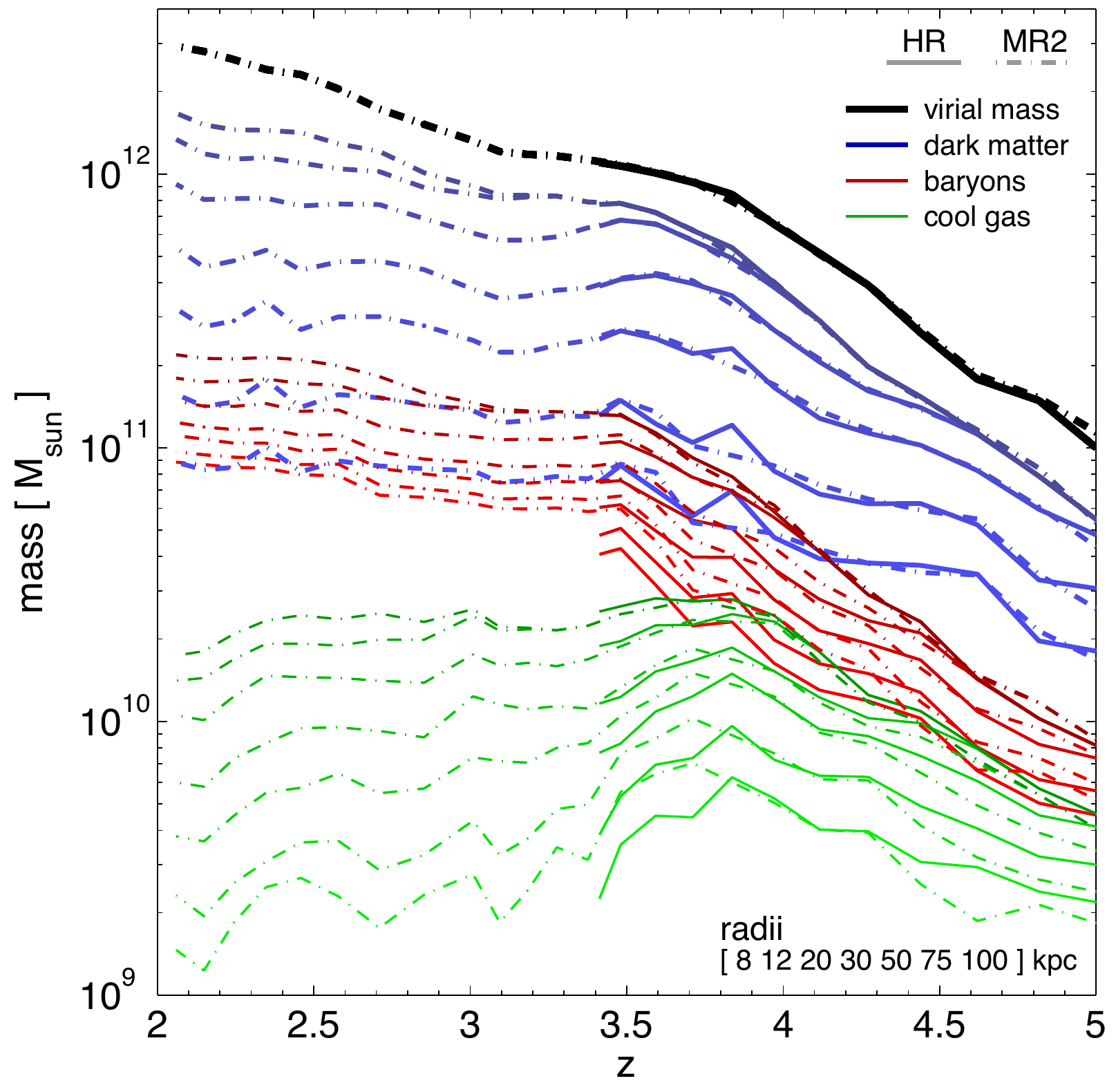} &
\includegraphics[width=80mm]{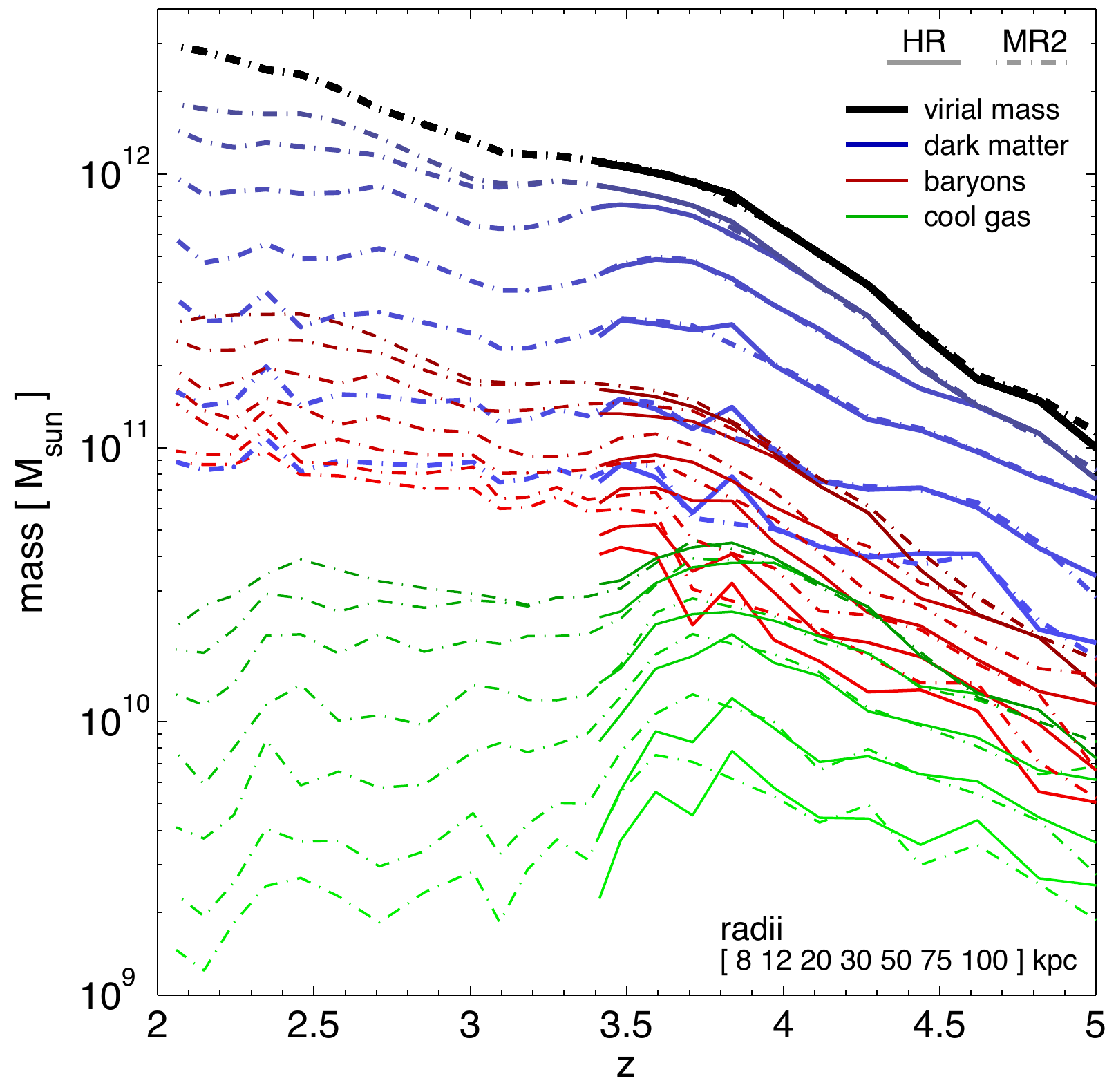} \\
\end{tabular}
\caption{Evolution of dark matter and baryonic mass within fixed proper radii. (Left panel) Only the contribution from the central galaxy and its halo are included. Satellites galaxies and sub-halos are removed. (Right panel) Same as left panel but including satellites and sub-halos. In each panel red curves show the baryonic (stellar and gas) mass within fixed proper radii ranging from 8 kpc (lower line) to min(R$_{\rm vir}$, 100 kpc) (upper line) as function of redshift. Blue lines show the evolution of the dark matter mass in the same fixed proper radii. Black lines indicate  the virial mass of the halo. The virial mass increases continuously because of accretion at large radii ($\gtrsim{}100$ kpc) and the drop of the background density of the Universe with cosmic time. In contrast, dark matter and baryonic mass in \emph{fixed proper radii} grow strongly only at $z>3.5$. At $z=2-3.5$, the growth of dark matter and baryons stalls.}
\label{fig:massInPhysicalRadii}
\end{figure*}

Furthermore, the reduction of the gas accretion and the subsequent drop in the SFR is not merely a consequence of hydrodynamics and gas cooling, but related to gravitational processes, as we show now. Figure~\ref{fig:massInPhysicalRadii} plots the evolution of the virial mass of the halo as well as the evolution of the mass of dark matter, baryons (gas and stellar matter), and cool gas enclosed in \emph{fixed proper radii}.

The dark matter mass, the baryonic mass, the cool gas mass, and the virial mass all grow at $z\gtrsim{}3.7$. The dark matter mass and the cool gas mass increase roughly in look-step, while the baryonic mass and the virial mass grow somewhat faster. Hence, at $z\gtrsim{}3.7$ the halo is in a phase of fast accretion (\emph{``collapse phase''}). Note that the mass increases within any fixed proper radius $r<R_{\rm vir}$ indicating genuine growth on all scales. During the collapse phase, the gas accretion rates and the SFR increase approximately exponentially and much of the in-situ formed stellar mass of the galaxy is built. 

At $z\lesssim{}3.5$ this phase of fast accretion comes to a halt. This can be seen clearly in the evolution of the dark matter, baryonic, and cool gas masses. For instance, the dark matter mass enclosed in a given physical radius up to $\sim{}75$ kpc does not evolve between $z=3.5$ and $z=2$. The figure shows that dark matter growth is truncated from outside in consistent with a shut-down of accretion starting on large scales. Specifically, the dark matter mass with 50 kpc increases until at $z=3.65$, while mass within 8 kpc grows until $z=3.45$. The transition from a fast to a slow accretion phase also affects the evolution of the baryonic masses, the cool gas mass, and (omitted from Fig.~\ref{fig:massInPhysicalRadii} for reasons of presentational clarity) the stellar mass. 

With the onset of this \emph{``cosmological starvation''} phase the growth of the baryonic (and stellar mass) slows down substantially. It does not completely stop, however, because the accretion of cold gas from large radii supports a small, but non-negligible amount of star formation. In addition, mergers increase the stellar mass with ex-situ formed stars.  The mass of cool gas decreases slowly at $z<3.7$, indicating that gas consumed in star formation is largely replenished by newly accreted gas. There is a marked difference in the cool gas-to-stellar mass ratio during the collapse phase ($\sim{}1$ at $z=4-5$) and in the starvation phase ($<0.1$ at $z=3.4-2$), see Table~\ref{tab:Results}.

\begin{figure}
\begin{tabular}{c}
\includegraphics[width=80mm]{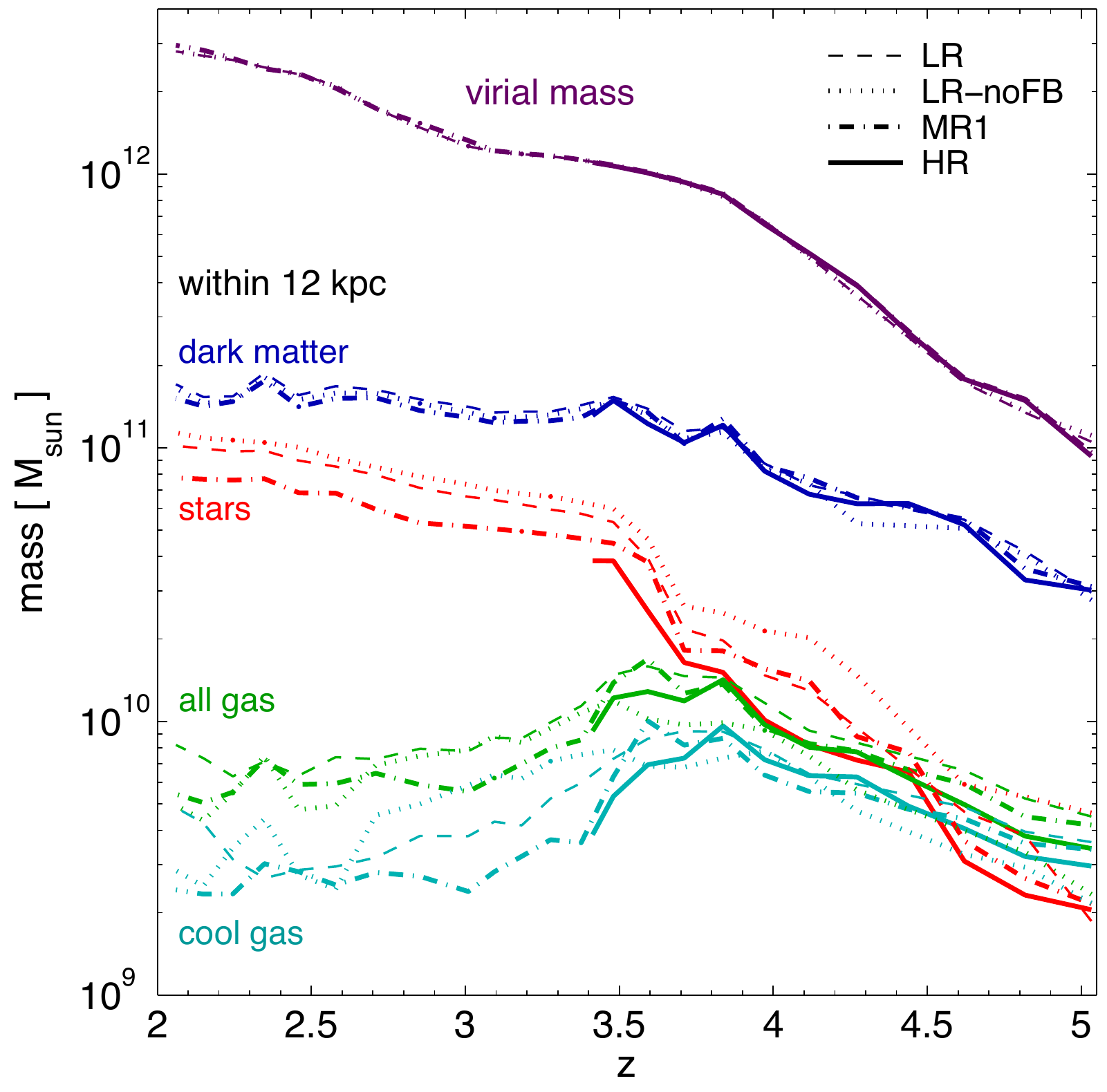}
\end{tabular}
\caption{Evolution of dark matter and baryonic masses within a sphere of 12 proper kpc radius around the central galaxy and evolution of the virial mass. We show (from bottom to top) the mass of cool gas ($T<2.5\times{}10^5$ K), the gas mass, the stellar mass, the dark matter mass (all within 12 kpc) as well as the virial mass of the halo (see legend). The dashed (dotted, dot-dashed, solid) lines show the mass components for the LR (LR-noFB, MR1, HR) run. The various mass components evolve in a similar way in the different runs suggesting that, on a qualitative level, our conclusions about cosmological starvation are neither affected by numerical resolution nor by the presence or absence of supernova feedback.}
\label{fig:massInPhysicalRadiiComponents}
\end{figure}

Cosmological starvation is not related to feedback from galaxies as we pointed out in section \ref{sect:LeavingSFS}. We prove this again in Fig.~\ref{fig:massInPhysicalRadiiComponents} which compares the evolution of the dark matter, stellar, and gas masses for the LR and LR-noFB runs. As expected, the no feedback run results in a somewhat larger stellar mass at any given time. However, aside from this change in the normalization the mass components evolve in the same way and show the same slow-down of their growth (or even decline) after $z\sim{}3.5$.

We should point out that the virial mass continues to grow in the cosmological starvation phase. The increase in virial mass in the absence of physical accretion, sometimes called ``pseudo-evolution'', is a consequence of tying the definition of virial mass to the evolving mean matter density (or the critical density) of the Universe \citep{2007ApJ...667..859D, 2013ApJ...766...25D}. Hence, the virial mass does not robustly indicate whether a galaxy transits from the collapse phase to the starvation phase. Instead, the dark matter mass or the baryonic mass within fixed proper radii should be used.

\section{Caveats}
\label{sect:Caveats}

\boldtext{In this study we analyze the evolution of a single massive galaxy at $z\sim{}2-5$.} The immediate question arises whether we can generalize our results based on a single object. In particular, it would be important to know whether our findings are representative of massive galaxies at high redshifts, or whether we accidentally picked a peculiar case. We selected the simulated galaxy based on its $z=0$ halo mass and on the condition of residing in a ``typical'' large scale environment (based on the matter over-density within 7 Mpc). Its halo accretion history differs from a purely exponential growth \citep{2002ApJ...568...52W}, as it grows quickly at early times and and slows down at later times. This is nothing special, though, as $\sim{}40\%$ of halos show such a behavior \citep{2009MNRAS.398.1858M}. Furthermore, we show that the simulated galaxy is a reasonably good representation of quiescent massive galaxies observed at $z=2$. Finally, we note that not all galaxies at high redshift are quiescent. In fact, roughly 20-50\% of massive galaxies are quiescent at $z=2$ and the fraction decreases toward higher redshift (e.g., \citealt{2013ApJ...777...18M}). These results are consistent with a cosmological starvation origin of massive quiescent galaxies at high redshift, i.e., massive galaxies are quiescent at $z=2$ if they reside in halos with a particularly early halo formation time. Clearly, future simulations with larger galaxy samples will be helpful in challenging or confirming this picture.

By comparing the LR, MR and HR runs of the Argo simulation we can test whether the results are numerical converged. We find that global galaxy properties, such as stellar masses, gas masses, or gas inflow rates show only small trends with numerical resolution. The most noticeable differences are the sizes of galaxies at $z>3$ and the entropy profile of hot gas around galaxies. The differences in sizes largely disappear by $z=3$. Outside the very central region of the halo the entropy profiles of the HR and MR runs are in agreement. In contrast, the LR run produces hot circum-galactic gas with a significantly lower entropy, which reduces the gas cooling time and increases the star formation activity. Analogous trends have been noted before in cosmological simulations of disk galaxies (e.g., \citealt{2008ASL.....1....7M}) and have been accredited to a reduced efficiency of supernova feedback at low numerical resolution. Specifically, low resolution models struggle at capturing the stochastic, bursty, and inhomogeneous injection of thermal energy by supernova explosions into the ISM thereby reducing the feedback efficiency. Overall, the agreement between the different runs is good (galaxy properties differ typically by less than 0.3 dex), especially at $z=2-3$, indicating reasonable convergence.

The simulated galaxy leaves the star forming sequence at $z\sim{}3.5$ and reduces its SFR to a few $M_\odot$ yr$^{-1}$ outside the central 300 pc (Figure~\ref{fig:SFhist}). In addition, the galaxy forms stars at a rate of $\sim{}10$ $M_\odot$ yr$^{-1}$ in its very innermost (and numerical unresolved) region. Measuring the SFR of massive, quiescent (according to the UVJ classification) galaxies observed at high redshift is challenging \citep{2014ApJ...783L..30U, 2014arXiv1402.0006H}. However, careful modeling of the spectral energy distributions (SEDs) of such galaxies shows that their SFRs are probably a few $M_\odot$ yr$^{-1}$ or less (e.g., \citealt{2013ApJ...771...85V}; consistent with the non-central SFR of the simulated galaxy) although some ``quiescent galaxies'' show SFRs up to $\sim{}30$ $M_\odot$ yr$^{-1}$ (e.g., \citealt{2014ApJ...783L..14S}). Nonetheless, SFRs are probably less than a few $\sim{}$ $M_\odot$ yr$^{-1}$ in at least some observed high redshift galaxies. It is unclear whether the physics implemented in our simulation is sufficient to reproduce massive galaxies with such low SFRs. We speculate that various AGN feedback channels (e.g., \citealt{2005MNRAS.358..168S, 2006MNRAS.365...11C}) may play a role here acting \emph{after} galaxies have already left the star forming sequence. Possibly they are only effective after accretion of cool gas onto the galaxy has started to decline. Subsequently, they may provide a path to shut down star formation completely.

\begin{figure*}
\begin{tabular}{cc}
\includegraphics[width=78.5mm]{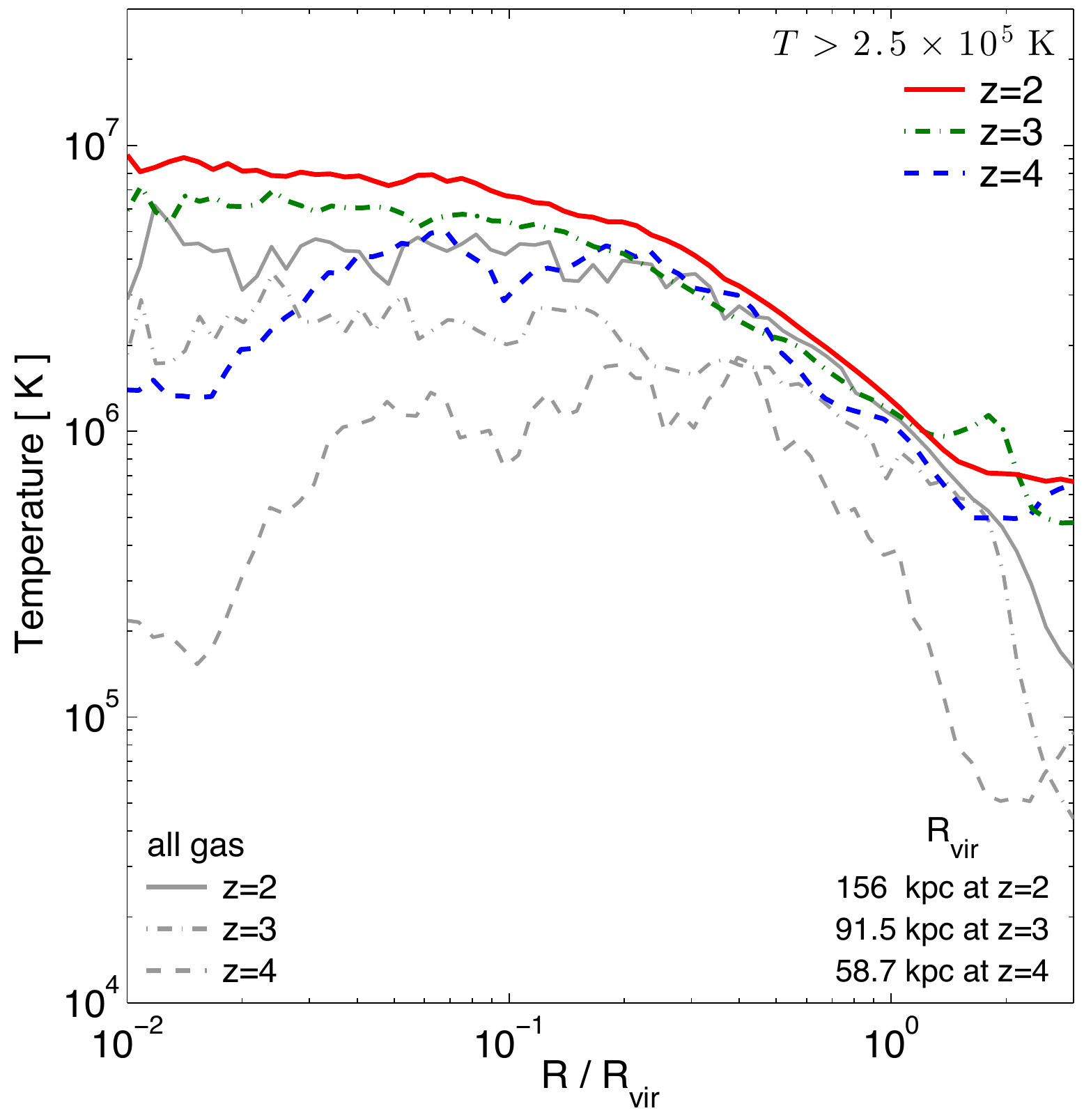} &
\includegraphics[width=80mm]{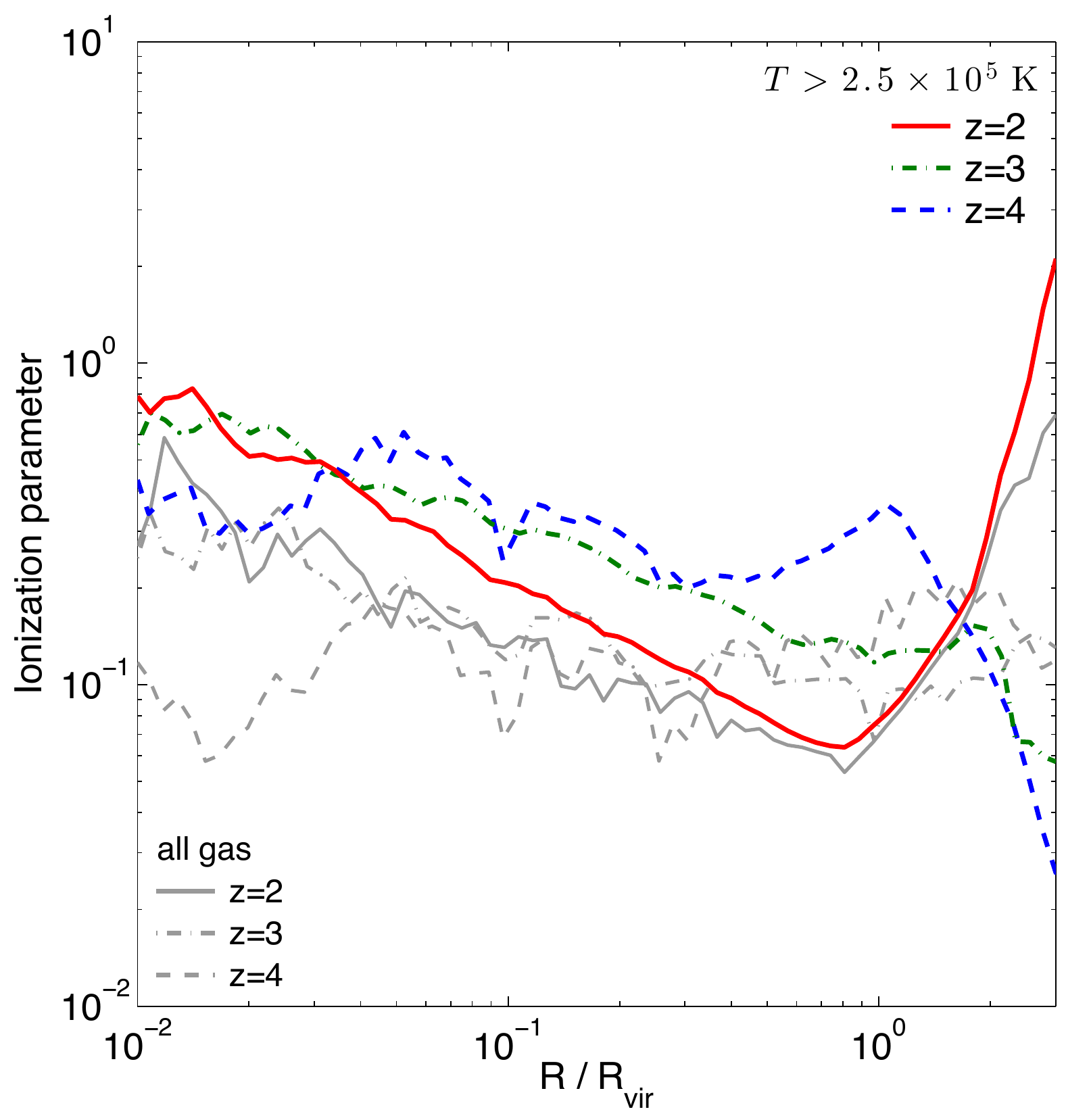} \\
\end{tabular}
\caption{ (Left) Average temperature vs distance from the center of the galaxy. (Right) Ionization parameter vs distance from the center of the galaxy. Plots are based on the MR2 run over the redshift range $z=2-4$. Colored lines show the temperature and the ionization parameter of the hot circum-galactic gas ($T>2.5\times{}10^5$ K). Gray curves show the temperature and the ionization parameter if no temperature cut is imposed. Solid (dot-dashed, dashed) lines show the results for $z=2$ ($z=3$, $z=4$). The hot circum-galactic gas has average temperatures of about $\sim{}2-8\times{}10^{6}$ K out to a large fraction of the virial radius. The ionization parameter of the circum-galactic gas is $0.1-0.8$.}
\label{fig:TvsR}
\end{figure*}

In this study we took a minimalistic approach by including only physical processes that we understand reasonably well (hydrodynamics, gravity) or that we add to match observed properties of galaxies (supernova feedback). However, we demonstrated that the choice of the feedback physics is not essential to explain why galaxies leave the star forming sequence at high redshift. Specifically, we found no qualitative changes of our results when we turn off feedback processes altogether.

We should point out that the Argo simulation accounts for radiative cooling based on a primordial gas composition, but does not include cooling via metal lines. We justify this simplification of the physical realism of our simulations as follows. First, given the high temperatures (a few $10^6$ K) and low metallicities ($Z\sim{}0.1$ $Z_\odot$) of the gas surrounding the simulated massive galaxy, we expect only moderate changes of the cooling rate (by about a factor 2 or less, see below). Such a change will not affect our findings on a qualitative level in agreement with conclusions reached by previous work (e.g., \citealt{2014MNRAS.438.3490T}). Second, uncertainties related to other aspects of our physical model (e.g., the modeling of feedback) largely outweigh the moderate quantitative changes that result from including or omitting metal line cooling. Third, gas cooling is also affected by UV and soft X-ray emission from nearby massive stars (.e.g, \citealt{2010MNRAS.403L..16C, 2014MNRAS.437.2882K}) and it is far from clear whether adding individual processes, e.g., metal line cooling, without their potential counterparts, e.g., radiative ionization by local sources, results in a better approximation of reality. Finally, we decided to mimic both the resolution and the physical model (which lacks high temperature metal line cooling) of the Eris simulation \citep{2011ApJ...742...76G} to extend its results to a higher mass regime.

In the left panel of Figure~\ref{fig:TvsR} we show the average gas temperature as function of distance from the galaxy at redshift 2, 3, and 4. The hot phase has a temperature of about $2-8\times{}10^{6}$ K out to a large fraction of the virial radius. The metallicity of the circum-galactic medium is about 0.1 $Z_\odot$ at $z\geq{}2$. Everything else being equal, metal line radiation increases the cooling rate by a factor 2 at $T\sim{}3.5\times{}10^6$ K compared with the cooling rate of gas with a primordial composition \citep{1993ApJS...88..253S}. However, adding metal line cooling in this straightforward way likely results in an overestimate of the cooling rate as we demonstrate now.

The right panel of Figure~\ref{fig:TvsR} shows the ionization parameter\footnote{The ionization parameter is defined as the number density of ionizing photons divided by the hydrogen density. We compute it from Eq. (1) of \cite{2010MNRAS.403L..16C} using the SFR and the gas density profile of the simulated galaxy.} as function of distance. The ionization parameter in the circum-galactic gas  is $\sim{}0.1-0.8$ out to $R_{\rm vir}$ at $z=2-4$. Photoionization by local sources (in addition to photoionization by a uniform UV background already included in our simulations) should both reduce the cooling efficiency of hot gas and heat the gas. An ionization parameter of $\sim{}0.1$ can suppress the net cooling by almost a factor 2 at $T\sim{}2\times{}10^6$ K \citep{2010MNRAS.403L..16C} and by much larger factors at lower gas temperatures (e.g., \citealt{2009MNRAS.393...99W}). In addition, ionization fractions may be out of equilibrium, which further reduces the cooling rate at $T<10^6$ K \citep{2007ApJS..168..213G, 2013MNRAS.434.1043O}. Finally, the radiation field of an AGN (not included in our simulation) could also heat the gas (via photo-ionization and Compton heating) and reduce the cooling \citep{2007ApJ...665.1038C, 2012ApJS..202...13G}.

While our main findings should be robust against moderate changes in the cooling rates, we acknowledge the general need for a more sophisticated modeling of gas cooling in numerical simulations. In particular, local radiation fields (see \citealt{2014MNRAS.437.2882K} for a first study along these lines) and non-equilibrium effects \citep{2013MNRAS.434.1043O} clearly matter, yet they are typically ignored in cosmological runs. A proper modeling of the cooling rates may also lower the amount of feedback required to offset excessive overcooling of gas. 

\section{Discussion}
\label{sect:Discussion}

In section \ref{section:OriginQuenching} we identified the leveling off and subsequent decrease of the gas accretion rate onto galaxies (cosmological starvation) as the trigger of star formation quenching in massive high redshift galaxies. Based on our results we argued that major galaxy mergers and subsequent AGN driven gas outflows, often suggested as the cause of star formation quenching (e.g., \citealt{2005ApJ...620L..79S, 2005Natur.433..604D}), \boldtext{are likely of secondary importance. Such processes may however expedite the transition from the star forming to the quiescent population by reducing the gas reservoir of galaxies via central starbursts and strong AGN driven outflows.} However, as Figure~\ref{fig:SFhist} shows, the suppression of the SFR can be quite fast ($\tau\sim{}100$ Myr) even in the case without AGN feedback or the occurrence of a major merger.

The process that quenched star formation and let to quiescent, early type galaxies in the local Universe probably lasted no longer than a few hundred Myr (e.g., \citealt{2014MNRAS.tmp..527S}). In contrast, the typical gas depletion time of a  ``normal'' star forming, nearby spiral is a few Gyr (e.g., \citealt{2008AJ....136.2846B}). The difference in time scales has been used to argue that AGN feedback is thus required to suppress star formation on sufficiently fast time scales \citep{2011MNRAS.415.3798K}. Here we demonstrate, however, that such a rapid decline in the SFR is also possible in compact, massive galaxies (which have short dynamical times) with the help of supernova feedback.

A different timescale constraint can be derived from the abundance ratio of alpha elements and iron in stellar atmospheres. It is typically understood that this ratio measures the time scale of star formation. Unfortunately, the observed ratio is also sensitive to the IMF, to the time delay function of type Ia supernovae, to the adopted stellar yields, and to metal losses via stellar winds, which complicates its interpretation. Star formation timescales derived this way typically range from $\sim{}0.2-0.8$ Gyr (see \citealt{2014ApJ...780...33C} and references therein), up to $\sim{}1$ Gyr \citep{1999MNRAS.302..537T}, but are broadly consistent with our findings (few hundred Myr).

We showed in Figure~\ref{fig:accretion rates} that the gas accretion rate shuts down before the SFR. Hence, even after cosmological starvation sets in, galaxies may continue their star formation and AGN activity for up to a few $\sim{}100$ Myr. If feedback from the AGN does not affect the dynamics of the ISM, galaxies likely shut down their activity from outside in, i.e., a declining SFR should precede a declining AGN activity. In this case galaxies on the star forming sequence as well as recently quenched galaxies may show AGN activity.

An idea often associated with the suppression of star formation in massive halos is the concept of a threshold halo mass \citep{2003MNRAS.345..349B, 2006MNRAS.368....2D, 2006MNRAS.370.1651C, 2010ApJ...718.1001B}. The threshold mass arises from the competition of gas cooling and gas compression in the shock-heated circum-galactic medium. The cooling time in massive halos is sufficiently long to support a stable virial shock that heats most of the accreted gas to the virial temperature of the halo. Additional heating sources such as radio mode AGN feedback or gravitational heating may further reduce the already low cooling rates of this shock-heated circum-galactic material.

However, numerical simulations show that at $z\geq{}2$ dense filaments of cool gas can penetrate the virial radius of even massive halos unshocked \citep{2005MNRAS.363....2K, 2009MNRAS.395..160K, 2008MNRAS.390.1326O, 2013MNRAS.429.3353N}. In fact, much of the star formation activity at high redshift is likely supported by inflowing streams of cool gas \citep{2009Natur.457..451D}. Hence, the threshold mass picture may explain the origin of massive, quiescent galaxies at $z<2$ \citep{2006MNRAS.370.1651C}, but it does not account for the large number of quiescent galaxies observed at $z\geq{}2$. Cosmological starvation is a complementary process that solves this problem by reducing the large scale gas accretion rate onto selected halos at high redshift. It is primarily a gravitational, not a hydrodynamical, process and affects all massive, high redshift galaxies that reside in halos undergoing a period of reduced growth.

Cosmological starvation also differs from the two-phase model of galaxy formation developed to explain the size growth of massive galaxies \citep{2007ApJ...658..710N, 2009ApJ...699L.178N, 2010ApJ...709..218F, 2010ApJ...725.2312O, 2012ApJ...744...63O}. The two phase model consists of an early dissipative phase in which much of the stellar mass of the central galaxy is formed resulting in a compact massive galaxy at high redshift. During the second phase of the two-phase model the (by then) quiescent galaxy grows in mass and size by mergers, i.e., via the accretion of stellar material. In contrast, the cosmological starvation phase is a period marked by little accretion of both gaseous and stellar matter. It is responsible for the abundance of massive, compact, quiescent galaxies at high redshift.

The picture we propose in this work is consistent with several previous findings. N-body simulations of dark matter often show a transition from a period of genuine growth to a period of starvation (e.g., \citealt{2007ApJ...667..859D}). During starvation the mass within fixed proper radii remains constant, while the virial mass still grows (``pseudo-evolution''; \citealt{2013ApJ...766...25D}). However, the importance of the starvation phase for the evolution of galaxies has not yet been fully appreciated, despite the close correlation between dark matter accretion and gas accretion onto halos \citep{2011MNRAS.414.2458V, 2011MNRAS.417.2982F}.

Empirical models that correlate colors and star formation activity of galaxies at fixed stellar mass with the \emph{halo formation time} are successful in predicting the clustering of galaxies as a function of their color or star formation activity, at least for galaxies in the local universe \citep{2013MNRAS.435.1313H, 2013arXiv1310.6747H, 2014arXiv1403.1578W}. Cosmological starvation (potentially combined with the formation of stable virial shocks) is a promising candidate that may provide the physical basis for the main ansatz of these models. 

Models that attribute quenching to feedback, from stellar sources or via an AGN, face the challenge to explain why the quiescent and the star forming galaxy populations substantially overlap in stellar mass (at $10^{11}$ $M_\odot$) at high redshift (e.g., \citealt{2014A&A...561A..86M}). If star formation quenching is related to feedback from the galaxy, why do some galaxies quench and other, similarly massive galaxies, do not? This observation is also a challenge for models based solely on a halo mass threshold. In such models one expects quenching to occur as soon as galaxies exceed a critical stellar mass. In contrast, the cosmological starvation picture naturally accounts for the co-existence of star forming and quiescent galaxies of the same stellar (and halo) mass. Here, the star formation activity is based on the recent rate of gas accretion rate onto the halo and not just on the the halo mass.

We can only speculate why the importance of cosmological starvation has not been recognized before. First, our work is probably the first cosmological simulation that reproduces the observed global properties of a massive, high redshift galaxy. Previous simulations with insufficient resolution and/or inadequate physics might not have recognized the starvation phase because of overcooling. Also, excessively strong feedback could have masked the modulation of gas accretion. Second, previous work often focussed on \emph{average} accretion rates and SFRs for halos of a given mass (e.g., \citealt{2011MNRAS.414.2458V, 2011MNRAS.417.2982F}). However, the point of the starvation model is that it distinguishes between halos of the same mass but with a different accretion history. Also, the fraction of massive, quiescent galaxies is only $\sim{}20\%$ at $z\gtrsim{}3$ \citep{2013ApJ...777...18M}. Hence, the impact of this significant galaxy population can be missed if only average or median properties are studied. Furthermore, the pseudo-evolution of the halo mass \citep{2013ApJ...766...25D} may obfuscate the connection between star formation and dark matter accretion. Pseudo-evolution has so far been ignored in galaxy evolution studies based on dark matter merger trees. Finally, as many simulations still struggle in producing galaxies with realistic stellar masses and SFRs, less attention has been given to study exactly when and how galaxies leave the star forming sequence.

\section{Summary and Conclusions}
\label{sect:Summary}

One of the major unsolved problems in galaxy evolution is to identify the physical process, or the processes, regulating the transition of galaxies from the star forming to the quiescent population. The old ages of massive, quiescent galaxies in the local Universe place this quenching of star formation at an early time, likely before $z\sim{}2$ \citep{2010MNRAS.404.1775T}. Similarly, the ages of massive, quiescent galaxies at $z=2-4$ (e.g., \citealt{2013ApJ...770L..39W, 2014ApJ...783L..14S}) indicate that some galaxies reduce their SFRs to low levels at even higher redshift. In this paper we analyze the origin of star formation quenching in high redshift galaxies with the help of a state-of-the-art cosmological simulation of a massive galaxy. 

Our main findings are as follows.
\begin{itemize}
\item The global properties of the simulated galaxy are in good agreement with those of similarly massive galaxies observed at high redshift. Reproduced properties include the stellar-to-virial mass ratio, the size of the stellar component, and the sSFR while on the star forming sequence.
\item The simulation includes thermal supernova feedback, but does not model feedback from AGN. This suggests that AGN feedback is not an essential ingredient to reproduce properties of massive, quiescent galaxies  at high redshift.
\item At $z\sim{}3.5$ the simulated galaxy leaves the star forming sequence and the sSFRs decrease by almost an order of magnitude within a few 100 Myr. The SFR declines approximately exponentially with a e-fold time of $\sim{}$ 100 Myr. By $z\sim{}2$ the colors and the stellar mass of the simulated galaxy agree with those of massive, quiescent galaxies present at those redshifts.
\item The drop of the sSFR is not caused by feedback processes, but rather a consequence of a leveling off and subsequent decline in the cool gas accretion rate onto the halo of the galaxy. 
\item The decrease of the sSFR is somewhat faster and more pronounced in our higher resolution runs compared with the low resolution run. We attribute this difference to the diminished efficacy of the implemented stellar feedback scheme at low numerical resolution. Hence, feedback likely plays a crucial role in suppressing star formation to the very low levels (less than a few $M_\odot$ yr$^{-1}$) observed in a large fraction of massive, quiescent, high redshift galaxies.
\end{itemize}

Based on our findings we propose a novel picture for the suppression of star formation that differs from previous suggestions based on a halo mass threshold or on star formation quenching via feedback-driven outflows during major mergers. After a period of fast gas accretion and exponential growing SFRs, some massive galaxies at high redshift enter a period of cosmological starvation in which the gas and dark matter accretion rates onto their halos first stall and subsequently decrease. Affected galaxies leave the main sequence as the stalled gas accretion rates no longer support exponentially increasing SFRs. 

The cosmological starvation picture physically connects the sSFR of high redshift galaxies to the gas accretion rate onto their halos. It makes a number of testable predictions. 

Generally, we expect that central galaxies residing in unrelaxed, still collapsing large scale structures have larger sSFRs than central galaxies embedded in a relaxed, virialized environment. Specifically, we predict that at $z\geq{}2$ central galaxies of a given stellar mass are more likely to be still star forming if they reside in a higher density environments, e.g., are surrounded by a larger number of satellite galaxies. In fact, this reversal of the local star forming -- density relation, namely larger \emph{specific} star formation rates in denser regions, has been observed at $z\sim{}1$ \citep{2007A&A...468...33E}. 

Also, we expect the region of shock-heated gas to expand outward during a period of reduced gas accretion rate and, thus, lower density of the pre-shocked infalling gas (see equation 29 in \citealt{2003MNRAS.345..349B}). Hence, cosmological starvation might also be tested by inferring and comparing the size of the virial shock around massive, quiescent galaxies and around star forming galaxies of the same mass. At $z<1$ environmental processes related to the presence of a hot, dilute atmosphere of shock-heated gas affect satellite galaxies and help in suppressing their SFR (e.g., \citealt{2014MNRAS.438..717K}). Satellite galaxies around massive, high redshift galaxies may thus potentially be used to probe the extent of the virial shock.

Finally, during the starvation phase the accretion of both gas and stellar material is reduced. Hence, we predict that recently quenched, high redshift galaxies have (on average) a different distribution of satellite galaxies (e.g., fewer satellites at large distances, fewer massive satellites) than star forming galaxies of the same stellar mass.

Cosmological starvation highlights the importance of the recent halo accretion history for the evolution of galaxies. Abundance matching techniques show that the stellar mass of galaxies is primarily controlled by their halo mass. Here we demonstrate that the accretion rate onto halos acts as an additional lever that controls the star formation rate of central galaxies at high redshift. Accretion rates and masses are tightly coupled for halos with an average accretion history \citep{2008MNRAS.383..615N}, but differ for the large fraction of halos that deviate from pure exponential growth \citep{2009MNRAS.398.1858M}. The exploration of the large variety of halo accretion histories in future work may lead to a deeper understanding of the observed diversity of high redshift galaxies.

\acknowledgements

We thank A. Babul, G. Barro, M. Kriek, P. Madau, D. Martizzi, D. Marchesini, J. X. Prochaska, E. Quataert,  E. Scannapieco, and F. van de Voort for inspiring discussions and valuable input. We are also grateful to the Aspen Center for Physics for organizing the winter 2014 conference ``Unveiling the Formation of Massive Galaxies - Theoretical and Observational Challenges'', which gave us the opportunity to present the results of this work to, and to receive stimulating suggestions from, the galaxy evolution community. We thank the Fermi National Accelerator Laboratory for supercomputing time. We also thank the Swiss Supercomputing Center for granting us early user access to the new Cray XC30 ``Piz Daint'' on which the HR run is being carried out. RF acknowledges support for this work by NASA through Hubble Fellowship grant HF-51304.01-A awarded by the Space Telescope Science Institute, which is operated by the Association of Universities for Research in Astronomy, Inc., for NASA, under contract NAS 5-26555. This work made extensive use of the NASA Astrophysics Data System and arXiv.org preprint server.\\


\appendix

\section{Functional form of the star formation history}
\label{sect:FuncFormSFH}

We can motivate exponentially growing and declining star formation histories (eq. \ref{eq:SFRexponential}) based on the conservation of gas mass and the ansatz of a linear relation between SFR and gas mass. The latter ansatz is not strictly accurate for the Argo simulation, as it uses a non-linear relation between SFR density and gas density, see section \ref{sect:Sim}. In fact, this non-linear relation implies that the SFR depends not only on gas mass, but also on the density distribution of the gas. The following ``derivation'' of eq. (\ref{eq:SFRexponential}) ignores these complications.

The mass of gas, $M_{\rm g}$, available for star formation in a galaxy changes because of gas accretion ($\dot{M}_{\rm g, acc}$), star formation (${\rm SFR}=M_{\rm g}/\tau_{\rm dep}$), galactic outflows (mass loading factor $\epsilon_{\rm out}$), and stellar mass loss (return fraction $R$), i.e.,
\begin{equation}
\dot{M}_{\rm gas}= \dot{M}_{\rm gas, acc} - \frac{1 - R + \epsilon_{\rm out}}{\tau_{\rm dep}}M_{\rm gas}.
\label{eq:MgConservation}
\end{equation}

We can easily solve this differential equation if we assume that $\hat{\tau}_{\rm dep} = \frac{\tau_{\rm dep}}{1 - R + \epsilon_{\rm out}}$ is a constant, and that $\dot{M}_{\rm gas, acc}$ depends on time, but not on $M_{\rm gas}$ itself. Basic calculus shows that
 the differential equation $\dot{x} + \beta{}x = \gamma(t)$ has the general solution $x(t) = (\Gamma(t) + C)e^{-\beta{}t}$ with $\Gamma(t)= \int{}^t\gamma(t')e^{\beta{}t'}dt'$ and constant $C$.

We now assume that there are two phases in the life a galaxy. A collapse phase in which gas accretion increases approximately exponentially with time $t$ since the big bang, i.e., $\dot{M}_{\rm g, acc}(t)=\gamma_{\rm G}\,e^{t/\tau_{\rm acc}}$ and a cosmological starvation phase in which the gas accretion levels off ($\dot{M}_{\rm g, acc}={\rm const}$) or stops altogether ($\dot{M}_{\rm g, acc}\approx{}0$). We further assume that the gas mass is zero at time $t=0$.

During the collapse phase the gas mass at time $t$ is
\begin{flalign*}
M_{\rm gas}(t) &= \frac{\gamma_{\rm G}}{\tau^{-1}_{\rm acc} + \hat{\tau}^{-1}_{\rm dep}}\left(e^{t/\tau_{\rm acc}} - e^{-t/\hat{\tau}_{\rm dep}}\right)\\
&\approx{}\frac{\gamma_{\rm G}}{\tau^{-1}_{\rm acc} + \hat{\tau}^{-1}_{\rm dep}}e^{t/\tau_{\rm acc}}\text{ if }t\gtrsim{}\tau_{\rm acc}.
\end{flalign*}

Hence,
\[
{\rm SFR} \approx{} \frac{\gamma_{\rm G}}{\tau_{\rm dep}/\tau_{\rm acc} + 1 - R + \epsilon_{\rm out}}e^{t/\tau_{\rm acc}}=\dot{M}_{\rm g, acc}\, \mathcal{O}(1),
\]
i.e., the SFR mirrors the gas accretion rate. The SFR is a pure exponential with zero offset during this phase.

For the starvation phase and the initial conditions ${\rm SFR}(t^*) = {\rm SFR}^*$, we obtain
\[
{\rm SFR} = O + ({\rm SFR}^* - O)e^{-(t-t^*)/\hat{\tau}_{\rm dep}},\text{ with }O = \frac{\dot{M}_{\rm g, acc}}{1 - R + \epsilon_{\rm out}}.
\]
Here $t^*$ is the time when the galaxy changes from the collapse phase into the starvation phase, i.e., the time when it leaves the star forming sequence. Physically, the SFR declines in the starvation phase on a gas depletion time scale because the gas reservoir is used up by star formation. The offset from a purely exponential decline is a consequence of continued gas accretion.

\end{document}